\documentstyle[12pt]{article}

\catcode`\@=11
\newdimen\@rotdimen
\newbox\@rotbox

\def\@vspec#1{\special{ps:#1}}
\def\@rotstart#1{\@vspec{gsave currentpoint currentpoint translate
   #1 neg exch neg exch translate}}
\def\@rotfinish{\@vspec{currentpoint grestore moveto}}
\def\@rotr#1{\@rotdimen=\ht#1\advance\@rotdimen by\dp#1%
   \hbox to\@rotdimen{\hskip\ht#1\vbox to\wd#1{\@rotstart{90 rotate}%
   \box#1\vss}\hss}\@rotfinish}
\def\@rotl#1{\@rotdimen=\ht#1\advance\@rotdimen by\dp#1%
   \hbox to\@rotdimen{\vbox to\wd#1{\vskip\wd#1\@rotstart{270 rotate}%
   \box#1\vss}\hss}\@rotfinish}%
\def\@rotu#1{\@rotdimen=\ht#1\advance\@rotdimen by\dp#1%
   \hbox to\wd#1{\hskip\wd#1\vbox to\@rotdimen{\vskip\@rotdimen
   \@rotstart{-1 dup scale}\box#1\vss}\hss}\@rotfinish}%
\def\@rotf#1{\hbox to\wd#1{\hskip\wd#1\@rotstart{-1 1 scale}%
   \box#1\hss}\@rotfinish}%
\def\rotate{\@ifnextchar[{\@rotate}{\@rotate[l]}}
\def\@rotate[#1]#2{\setbox\@rotbox=\hbox{#2}\@nameuse{@rot#1}\@rotbox}

\catcode`\@=12

\def\NPB#1#2#3{Nucl. Phys. B{#1} (19#2) #3}
\def\PLB#1#2#3{Phys. Lett. B{#1} (19#2) #3}

\def\PRD#1#2#3{Phys. Rev. D{#1} (19#2) #3}

\def\ds{\displaystyle}

\def\yzero{\smash{\hbox{$y\kern-4pt\raise1pt\hbox{${}^\circ$}$}}}

\def\-{\hphantom{-}}
\def\ov{\overline}
\def\s2{\frac{1}{\sqrt2}}
\def\s22{\frac{1}{2\sqrt2}}

\def\oh{\frac{1}{2}}
\def\beq{\begin{equation}}
\def\eeq{\end{equation}}
\def\beqa{\begin{eqnarray}}
\def\eeqa{\end{eqnarray}}
\def\tr{{\rm tr \,}}
\def\Tr{{\rm Tr \,}}
\def\IF{\relax{\rm I\kern-.18em F}}
\def\II{\relax{\rm I\kern-.18em I}}
\def\IP{\relax{\rm I\kern-.18em P}}
\def\IC{\relax\hbox{\kern.25em$\inbar\kern-.3em{\rm C}$}}
\def\IR{\relax{\rm I\kern-.18em R}}

\def\Dsl{\,\raise.15ex\hbox{/}\mkern-13.5mu D} 
\def\IZ{Z\kern-.4em  Z}
\def\cp#1{\relax\ifmmode {\IP\kern-2pt{}_{#1}}\else $\IP\kern-2pt{}_{#1}$\fi}

\begin{document}

\makeatletter
\@addtoreset{equation}{section}
\makeatother
\renewcommand{\theequation}{\thesection.\arabic{equation}}
\pagestyle{empty}
\rightline{ FTUAM-97/9}
\rightline{\tt hep-th/9707075}
\begin{center}
\LARGE{
 D=6 , \  N=1  String Vacua and Duality
\footnote{To appear in the proceedings of the APCTP Winter School on Duality,
Mt. Sorak (Korea), 
February 1997.} 
\\[10mm] }
\large{
L.~E.~Ib\'a\~nez   and 
A.~M.~Uranga
\\[0.8in]}
\small{
 Departamento de F\'{\i}sica Te\'orica C-XI \\ and\\
Instituto de F\'{\i}sica Te\'orica  C-XVI, \\
 Universidad Aut\'onoma de Madrid,\\
Cantoblanco, 28049 Madrid, Spain. 
\\[1.5in]}
\small{\bf Abstract} \\[0.2in]
\end{center}

\begin{center}
\begin{minipage}[h]{5.5in}

We review the structure  $D=6$, $N=1$ string vacua with emphasis 
on the different connections due to $T$-dualities and $S$-dualities.
The topics discussed include:  Anomaly cancellation; 
K3 and orbifold  $D=6$, $N=1$ heterotic compactifications; 
$T$-dualities between  $E_8\times E_8$ and
$Spin(32)/Z_2$ heterotic vacua; non-perturbative heterotic vacua 
and small instantons; $N=2$ Type-II/Heterotic duality in $D=4$ ;
F-theory/heterotic duality in $D=6$; and heterotic/heterotic duality in 
six and four dimensions. 
\end{minipage}
\end{center}
\newpage

\setcounter{page}{1}
\pagestyle{plain}
\renewcommand{\thefootnote}{\arabic{footnote}}
\setcounter{footnote}{0}

\section{Introduction}

The last few years of work on non-perturbative
$S$-dualities  \cite{filq} have taught us a lot about the 
connections and equivalences  among different string 
vacua in different dimensions (for reviews, see ref.\cite{reviews}).
Also a good amount of non-perturbative aspects 
of string theories has been learnt.  Of course, one
would be particularly interested in the understanding
of  non-perturbative vacua in $D=4$  with  $N=1$ or $N=0$
supersymmetries.
However, 
as one goes to lower dimensions and  smaller number of 
supersymmetries,   the physics  becomes  more and more
non-trivial.  It is thus interesting to proceed step by step
and  try first to understand as much as possible of the 
dynamics of  theories  with higher number of dimensions
and/or  supersymmetries.  In this respect, six dimensional vacua 
with  the minimal number of supersymmetries, $N=1$, are  particularly
interesting because of several reasons: 1) These theories 
are chiral and
the cancellation of anomalies restricts the physics substantially.
One can  use this constraint both as a check  of the
consistency of the vacua and as a guide for the search of
new non-perturbative dynamics.  2) When toroidally compactified,
these  theories yield  4-dimensional vacua  with $N=2$  
supersymmetry, for which a number of non-perturbative 
results are known. 3)  One  expects   that some of the
non-perturbative physics going on for $D=6$, $N=1$ 
theories have a reflection in $D=4$. Thus, for example, 
one can consider 4-dimensional $N=1$ vacua
obtained upon  heterotic compactification on a CY manifold which is
a K3 fibration. When the size of the base is large
the theory looks locally like  the heterotic compactified on K3  and
some non-perturbative phenomena  (like e.g. small instanton effects)  
are  inherited from  known $D=6$  dynamics;
4) Some $D=6$ string  vacua have suggested the existence of 
new classes of non-trivial  renormalization group fixed points of $D=6$ 
field theory \cite{gh, sw6d,  dlp, seiberg}.  At
special points in the moduli space of  some of  the $D=6$, $N=1$ vacua 
  non-critical  tensionless strings  appear, associated to these new 
classes of non-trivial field theories. 

$D=6$, $N=1$   vacua  have been constructed
 in  essentially  three  ways: 1)  heterotic
compactifications  on a K3  manifold (or orbifold); 2)   F-theory compactified 
on elliptic  Calabi-Yau threefolds  ;  3) Type-IIB   $Z_N$  orientifolds.   
All these constructions   are related  and  one can often construct
the same model (possibly in different regions of the moduli space) by 
using different techniques. In these lectures  we will mostly discuss  
theories constructed using the first two.   The structure of these 
lectures is as follows.
In chapter 2, after reviewing the constraints coming from 
anomaly cancellation,  we describe  the construction of  
$D=6$, $N=1$  heterotic vacua  in terms of toroidal orbifolds and
K3 compactifications.  The $T$-dualities  among  $E_8\times E_8$  and
$Spin(32)/Z_2$   vacua are also discussed.   In section 2.3  
non-perturbative heterotic vacua and their connection with
the physics of small instantons are summarized.
In chapter 3 we  first discuss the structure of $D=4$, $N=2$  
string vacua. They are of interest for the purposes of these lectures 
since these vacua appear e.g. after 
trivial toroidal compactification of $D=6$, $N=1$ theories.  We then  
describe 
the dualities between Type-IIA  theory compactified on   a Calabi-Yau and 
heterotic 
compactified  on $K3\times T^2$.  This is better understood from the
perspective of the $D=6$  duality between F-theory  and heterotic string 
\cite{fth, mv1, mv2}
which is discussed in some detail in the rest of the chapter.   
In  the last chapter  we discuss heterotic/heterotic duality  
\cite{Dufflu, Duffkhuri, Minasian, Duff, dmw, afiq2}
 from different perspectives
including  usual heterotic K3 compactifications,  Type-IIB  orientifolds and 
F-theory.

\section{$D=6$, $N=1$ Heterotic Vacua}

\subsection{Gauge and  Gravitational Anomalies}

The  relevant supermultiplets in a $D=6$, $N=1$ theory are as follows:
\beqa
 & {\rm Gravity} & \rightarrow (3,3)\ + \ 2(2,3) \ +\  (1,3) \nonumber\\
 & {\rm Tensor} &  \rightarrow  (3,1) \ +\ 2(2,1) \ +\  (1,1) \nonumber\\
 & {\rm Vector} &  \rightarrow  (2,2) \ +\  2 (1, 2)    \nonumber\\
& {\rm Hypermultiplet} &    \rightarrow  2 (2,1) \ + \ 4 (1,1)
\label{multi6}  
\eeqa
where  the transformation properties with respect to the little group
$Spin(4) \simeq   SU(2)\times SU(2)$  group are shown.  Notice the following relevant facts:
1) There are scalar fields only in the hypermultiplets (two complex scalars) 
 and in the tensor multiplets (one real scalar). The vector multiplet does not contain 
scalars so there is no Coulomb phase associated to the vector multiplets 
in $D=6$. 
On the other hand there is a Coulomb phase associated to tensor multiplets  \cite{sw6d} 
since they contain a scalar  and,  upon reduction to lower dimensions, a
vector boson appears from the two index antisymmetric field.  2)  The gravity
multiplet contains a self-dual two index antisymmetric field  $B_{\mu \nu }^+$
and the tensor multiplets contain anti-selfdual two-form fields   
$B_{\mu \nu }^-$.   In a theory with just one tensor multiplet
the $B_{\mu \nu }^-$  can combine with the $B_{\mu \nu }^+$ from the
gravity multiplet to form an unconstrained field, for which a local 
action may be written.  This is the case of perturbative compactifications
of the heterotic string down to  six dimensions for which a single tensor
multiplet is inherited  from the $D=10$ gravity multiplet.  
It is also the case for  {\it smooth}  K3 compactifications of Type-I theory.
 There is no known local Lagrangian description for theories with more than 
one tensor multiplet  although, as we will see later on, theories of that type
frequently appear in  heterotic non-perturbative string vacua. In Type-I 
theory, tensor multiplets appear even at the perturbative level in the 
presence of orbifold singularities in the compact manifold
\cite{gj, dabol1, jp}.
3)  Due to supersymmetry, the scalars in hypermultiplets and those in
tensor multiplets are decoupled. In particular the metric in the hypermultiplet
moduli space is independent of the tensors  and viceversa.  Also,  the kinetic terms 
of vector multiplets only depend on the tensor multiplets and not on the
hypermultiplets 
\cite{sagn}.  4)  At the perturbative level, the only dynamics 
present in this class of theories is that of the Higgs mechanism.  In this 
process  de number of hypermultiplets $n_H$  minus the number of vector 
multiplets $n_V$  remains constant,   $\Delta (n_H-n_V)=0$.
5) Upon further toroidal compactification to four dimensions,   $D=6$ 
tensor multiplets give rise to $D=4, N=2$ vector multiplets  whereas 
hypermultiplets and vector multiplets  remain being so. In addition,
 extra vector multiplets  (often named $T,U$ ) associated to the torus and
one containing the dilaton ($S$)  appear in the spectrum.   Notice that in $D=4$
there is a Coulomb phase associated to vector multiplets, since the 
latter
now contain scalars. This observation is  relevant when studying the 
duality between Type-IIA compactified in a  Calabi-Yau (CY)  and
the  heterotic string compactified on $K3\times T^2$.

Gauge and gravitational anomaly cancellation restricts very strongly
the possible dynamics in $D=6, N=1$ theories.  Cancellation of the 
pure $R^4$ gravitational anomalies requires \cite{gsw, walton}:
\beq
n_H\ -\ n_V\ =\ 273\ -\ 29 n_T
\label{grava}
\eeq
where $n_{H,V,T}$ are respectively the number of  hyper, vector and tensor 
multiplets.  Notice the fact that one tensor multiplet contributes as much as 
29 hypermultiplets to the gravitational anomalies. This turns out to play an
important role in non-perturbative transitions.  If there is
a simple gauge group $G_a$, cancellation of the pure $F_a^4$  anomaly requires
\cite{erler}  :
\beq
T_a\ =  \ \sum _i n_i t_a(R_i) 
\label{anomg}
\eeq
where $n_i$ denotes the number of hypermultiplets transforming in the 
$R_i$ representation, and the sum runs over the different representations. 
$T_a, t_i(R_i)$ are defined by 
\beqa
\Tr F_a^4\  & = & \ T_a\ \tr F_a^4 \ + \ U_a (\tr F_a^2)^2 \nonumber \\
\tr F_a^4\  & =& \ t_a(R_i)\ \tr F_a^4 \ + \  u_a(R_i)  (\tr T_a^2)^2
\label{tT}
\eeqa
where Tr (tr) indicates trace in the adjoint (fundamental). 
Due to the absence of an independent fourth order Casimir, there are 
no quartic gauge anomalies for the exceptional groups and for $SU(2)$ 
and $SU(3)$.  For the classical groups SU(N), SO(N) and Sp(N), 
one has  \cite{erler} $T_a= 2N$, $(N-8)$ and $(N+8)$ respectively 
\footnote{Notice that the group
 $SO(8)$ is special: it is chiral in $D=6$ but anomaly free in the absence 
of  hypermultiplets.}.
Concerning $t_a$,  one has $t_a= 1$, $(N-8)$ and ${{\frac 12}(N^2-17N+54)}$ 
for the fundamental, 2-index and 3-index  antisymmetric representations
for all classical groups
\footnote{Notice that  the 2-index antisymmetric  representation 
is anomaly free for $SU(8)$, $SO(8)$ and $Sp(8)$.}.
For the spinorial representations of $SO(2M)$ groups one has
$t_a=-2^{(M-5)}$.
Notice that cancellation of gauge and gravitational anomalies are
consistent with the transitions:
\beqa
1\ {\rm tensor} \  & \leftrightarrow  & {\bf 28}\ + \ 1  \  
{\rm hypermultiplet} \nonumber \\ 
SO(8)\ + \  1\ {\rm tensor}   & \leftrightarrow  &  \ 1\ {\rm hypermultiplet}
\label{transia}
\eeqa
where the ${\bf 28}$ may be  any anomaly free representation
of dimension 28 like 
 e.g. the 2-index antisymmetric representation of
the $N=8$ classical groups, or  one half hypermultiplet in a {\bf 56} 
of $E_7$, etc.
It turns out that these kind of transitions are physically realized by
certain non-perturbative phenomena, as we will describe below.
Once the pure quartic gravitational and gauge anomalies cancel, the
anomaly polynomial $A_8$ which is left takes the form:
\beqa
A_8\ & =& \  (1-{\frac {n_T-1}8})(\tr R^2)^2\ -\ 
\tr R^2 \sum _a{\tilde {C_a}}\tr F_a^2\ +\   \nonumber\\
& + & \sum _a{\tilde {U_a}}(\tr F_a^2)^2\ + \ \sum _{a<b} Y_{ab} 
\tr F_a^2 \tr F_b^2 \label{polia}
\eeqa
where  the  values of the coefficients ${\tilde {C_a}}$, ${\tilde {U_a}}$ and
$Y_{ab}$  may be found in ref.\cite{erler}.  For our present purposes 
it is 
only necessary to recall that  for the case of a single tensor multiplet, 
$n_T=1$, one can rewrite  $A_8$ in the factorized form:
\beq
A_8\  = \  (\tr R^2-\sum _a V_a \tr F_a^2)(\tr R^2-\sum _a {\tilde {V_a}} 
\tr F_a^2) \label{factor}
\eeq
where $V_a$ is a gauge group factor which only depends on the group  and 
is equal to $2$, $1$, $1/3$, $1/3$, $1/6$ and $1/30$ for $SU(N)$, $SO(N)$, 
$F_4$, $E_6$, $E_7$ and $E_8$, respectively.  On the other hand, ${\tilde 
{V_a}}$ 
depends in addition on the representations of the hypermultiplets. In 
particular, one finds for example \cite{erler}  :
\beqa
{\tilde {V}}_{SU(N)}  &  = &  n_{a2}+ (N-4)n_{a3}+ {\frac 12}(N-4)(N-5)n_{a4}-2 
\ \  (for\ N\geq 4 ) \nonumber\\
{\tilde {V}}_{SO(2N)} &  = &  2^{(N-6)}n_s-2\  \ (for\  N\geq 3) 
\nonumber \\ {\tilde {V}}_{E_6} & =  &  {\frac 16}(n_{27}-6)\ ;\ 
{\tilde {V}}_{E_7} = {\frac 16}(n_{56}-4)\ ;\ 
{\tilde {V}}_{E_8} = {\frac  {-1}5}
\label{vtildes}
\eeqa
where $n_{ai}$, $i=2,3,4$ is the number of hypermultiplets in the $i$-index 
antisymmetric representation  and $n_s$ is the number of spinorial representations.
The $A_8$ anomaly above has  the appropriate form to be  cancelled 
by the exchange of the unconstrained antisymmetric field $B_{\mu \nu }$
present in the $n_T=1$ case.
Notice that these values of the ${\tilde {V_a}}$'s correspond to the case 
with just one
tensor multiplet. When several tensors are present  there is a modified 
factorization of  $A_8$,  with  ${\tilde {V_a}}$ depending on $n_T$  and 
an extra (perfect square) piece in $A_8$ which involves only
the gauge fields and is proportional to $(n_T-1)$. Then, a  generalized 
Green-Schwarz mechanism is at work \cite{sagn}, 
 in which this extra  piece is cancelled by the exchange
of  $(n_T-1)$ tensor multiplets.
If a $U(1)$ gauge theory is present, one has  for the group theory coefficients in
eq.(\ref{polia})  ${\tilde C}={\frac 16} \tr Q^2$ and ${\tilde U}={\frac 
23} \tr Q^4$ \cite{erler} . 
Now, standard factorization  as in eq.(\ref{factor})  is obtained only if
${\tilde U}={\tilde C}-1$.  It turns out that in plenty of $D=6$, $N=1$ vacua
this condition is not verified, even in perturbative models.  In those cases what
actually happens is that the $U(1)$ is spontaneously broken and swallows
one of the $B$-field modes with indices in the compact dimensions
\cite{gsw, berkooz, afiu}.

Due to supersymmetry, the anomaly coefficients $V_a$, ${\tilde V}_a$  are
related to the kinetic terms of vector multiplets. In particular, in the $n_T=1$ 
case one finds \cite{sagn} :
\beq
L_{gauge}^{D=6}\ =\
-\frac{(2\pi )^3}{8{\alpha }' } {\sqrt {G}}
\left (V_{\alpha }e^{-\phi /2} \ +\ {\tilde V}_{\alpha }e^{\phi /2}
\right)
\tr F_{\alpha }^2
\label{sagno}
\eeq
where  $\phi $ is the real scalar in the unique tensor multiplet present, i.e., the 
dilaton. This will turn out to be relevant when we discuss
heterotic/heterotic duality .

\subsection{Perturbative  Heterotic Vacua}

{\bf  i)  Toroidal $Z_N$ Orbifolds}

We  turn now to  consider   explicit $D=6, N=1$  heterotic vacua.
An interesting class of $D=6$, $N=1$ heterotic vacua can be obtained
from  symmetric toroidal orbifold compactifications on $T^4/Z_M$.
The construction of these models parallels that of $T^6/Z_M$ orbifolds
\cite{dhvw,orbi} as considered in refs.~\cite{walton, erler,afiq}. 
Here we
briefly review the notation and the salient points relevant to our discussion. 
Acting on the (complex) bosonic transverse coordinates,
the $Z_M$ twist $\theta$ has eigenvalues $e^{2 \pi i\, v_a}$,
where $v_a$ are the components of $v=(0,0,\frac1M,-\frac1M)$.
$M$ can take the values $M=2,3,4,6$. The embedding
of $\theta$ on the gauge degrees of freedom is usually
realized by a shift $V$  (not to be mistaken with the $V_a$ coefficients
that we introduced in the previous subsection!) 
such that $MV$ belongs to the $E_8\times E_8$ lattice 
$\Gamma_8 \times \Gamma_8$ or to the  $Spin(32)/Z_2$ lattice $\Gamma_{16}$.
This shift is restricted by the modular invariance constraint 
\beq
M\, (V^2-v^2)={\rm even}
\label{modinv}
\eeq
All possible embeddings for each $M$ are easily found.
In the $E_8\times E_8$ case, we find 2 inequivalent embeddings
for $Z_2$, 5 for $Z_3$, 12 for $Z_4$ and 59 for $Z_6$,
leading to different patterns of $E_8\times E_8$ 
breaking to rank 16 subgroups.
For $Spin(32)/Z_2$ we find 3 inequivalent embeddings
for $Z_2$, 5 for $Z_3$, 14 for $Z_4$ and 50 for $Z_6$.
Each of these models is only the starting point
of a bigger class of vacua, generated by adding Wilson
lines in the form of further shifts in the gauge
lattice satisfying extra modular invariance constraints,
by permutations of gauge factors, etc..

The spectrum for each model is subdivided in sectors. There are $M$ sectors 
twisted by $\theta^j$, $j=0,1,\cdots , M-1$. Each particle state is 
created by 
a product of left and right vertex operators  $L\otimes R$. At a generic point 
in the four-torus moduli space, the massless states follow from
\beq
\label{uno}
m_R^2=N_R+\frac{1}{2}\, (r+j\,v)^2
+E_n-\frac{1}{2} \quad ;\quad
m_L^2=N_L+\frac{1}{2}\, (P+j\,V)^2+E_j-1
\eeq
Here $r$ is an $SO(8)$ weight with $\sum_{i=1}^4 r_i=\rm odd$
and $P$ a  gauge lattice vector with $\sum_{I=1}^{16} P^I= \rm even$.
$E_j$ is the twisted oscillator contribution to the zero point energy and it 
is given by $E_j=j(M-j)/M^2$. The multiplicity of states satisfying 
eq.~(\ref{uno}) in a $\theta^j$ sector is given by the appropriate 
generalized GSO projections \cite{afiq}. In the untwisted sector there
appear the gravity multiplet, a tensor multiplet, charged hypermultiplets
and 2 neutral hypermultiplets (4 in the case of $Z_2$). In the twisted
sectors only charged hypermultiplets appear.
The generalized GSO projections are particularly simple in the $Z_2$
and $Z_3$ case since essentially all massless states survive with the same
multiplicity. 
The spectra for all $Z_2$ and 
$Z_3$ embeddings are shown in Tables~\ref{tabla1} and \ref{tabla2}
(from ref.\cite{afiuv}).

\begin{table}[p]
\footnotesize
\renewcommand{\arraystretch}{1.25}
\begin{center}
\begin{tabular}{|c|c|c|c|}
\hline
Shift $V$ & & & \\
\cline{1-1}
Group & \raisebox{2.5ex}[0cm][0cm]{Untwisted matter} &
\raisebox{2.5ex}[0cm][0cm]{Twisted matter} &
\raisebox{2.5ex}[0cm][0cm]{$(k_1,k_2)$} \\
\hline\hline
$\frac12(1,1,0, \cdots, 0)\times (0, \cdots, 0)$ &
(56,2)+4(1,1) &
8(56,1)+32(1,2)$^*$  & (24,0) \\
\cline{1-1}
$E_7\times SU(2)\times E_8$ & & & \\
\hline\hline
$\frac12(1,0, \cdots, 0)\times (1,1,0 \cdots, 0)$ &
 (1,56,2)+4(1,1,1) &
8(16,1,2) & (16,8) \\
\cline{1-1}
$SO(16)\times E_7\times SU(2)$ & + (128,1,1) &  & \\
\hline
\multicolumn{4}{c}{} \\[-0.5cm]
\hline
$\frac13(1,1,0, \cdots, 0)\times (0, \cdots, 0)$ &
(56,1)+3(1,1) &
9(56,1)+18(1,1)$^*$  & (24,0) \\
\cline{1-1}
$E_7\times U(1)\times E_8$ & & + 45 (1,1)$^*$ & \\
\hline\hline
$\frac13(2,0, \cdots, 0)\times \frac13(2,0 \cdots, 0)$ &
(14,1)+(64,1) + &
9(14,1)+9(1,14)  & (12,12) \\
\cline{1-1}
$SO(14)\times SO(14)\times U(1)^2$ & (1,14)+(1,64) + 2(1,1) & + 18(1,1)$^*$ & \\
\hline\hline
$\frac13(1,1,1,1,2,0,0,0)\times (0, \cdots, 0)$ &
(84,1)+2(1,1) &
9(36,1)+18(9,1)$^*$  & (24,0) \\
\cline{1-1}
$SU(9)\times E_8$ & & & \\
\hline\hline
$\frac13(1,1,2,0, \cdots, 0)\times \frac13(1,1,0 \cdots, 0)$ &
(27,3,1) + (1,1,56)   &
9(27,1,1)+9(1,3,1)  & (18,6) \\
\cline{1-1}
$E_6\times SU(3)\times E_7\times U(1)$ & + 3(1,1,1) & + 18(1,3,1)$^*$ & \\
\hline\hline
$\frac13(1,1,1,1,2,0 \cdots, 0)\times \frac13(1,1,2,0,0,0)$ &
(1,27,3) + (84,1,1)   &
9(9,1,3) & (15,9) \\
\cline{1-1}
$SU(9)\times E_6\times SU(3)$ & + 2(1,1,1) & & \\
\hline
\end{tabular}
\end{center}
\caption{Perturbative $Z_2$ and $Z_3$, $E_8\times E_8$, orbifold models.  The asterisk 
indicates twisted states involving left-handed oscillators. The last column 
shows which smooth $K3$ compactification yields a similar massless spectrum 
{\it after Higgsing}.
\label{tabla1} }
\end{table}

\begin{table}[p]
\footnotesize
\renewcommand{\arraystretch}{1.25}
\begin{center}
\begin{tabular}{|c|c|c|c|}
\hline
Shift $V$ & & & \\
\cline{1-1}
Group & \raisebox{2.5ex}[0cm][0cm]{Untwisted matter} &
\raisebox{2.5ex}[0cm][0cm]{Twisted matter} &
\raisebox{2.5ex}[0cm][0cm]{$G_0$} \\
\hline\hline
$\frac12(1,1,0, \cdots, 0)$ &
(28,2,2)+4(1,1,1) &
8(28,1,2)+32(1,2,1)$^*$  & $SO(8)$ \\
\cline{1-1}
$SO(28)\times SU(2)\times SU(2)$ & &  & \\
\hline\hline
$\frac12(1,1,1,1,1,1,0, \cdots, 0)$ &
(12,20)+4(1,1)  &
8(32,1) & $SO(8)$ \\
\cline{1-1}
$SO(12)\times SO(20)$ & &  &  \\
\hline\hline
$\frac14(1, \cdots, 1,-3)$ &
(120) + ($\ov{120}$)  &
8(16) + 8($\ov{16}$) & $1$ \\
\cline{1-1}
$SU(16)\times U(1)$ & + 4(1) &  &  \\
\hline
\multicolumn{4}{c}{} \\[-0.5cm]
\hline
$\frac13(1,1,0, \cdots, 0)$ &
(28,2)+3(1,1) &
9(28,2)+18(1,1)$^*$  & $SO(8)$ \\
\cline{1-1}
$SO(28)\times SU(2)\times U(1)$ & & + 45 (1,1)$^*$ & \\
\hline\hline
$\frac13(1,1,1,1,2,0, \cdots, 0)$ &
(22,5)+(1,10)  &
9(22,1)+9(1,10)  & $SO(8)$ \\
\cline{1-1}
$SO(22)\times SU(5)\times U(1)$ & + 2(1,1) & + 18(1,5)$^*$ &  \\
\hline\hline
$\frac13(1,1,1,1,1,1,1,1,0,\cdots, 0)$ &
(16,8)+(1,28) &
9(1,28)+18(1,1)$^*$  &  $SO(8)$\\
\cline{1-1}
$SO(16)\times SU(8) \times U(1)$ & + 2(1,1) & & \\
\hline\hline
$\frac13(1,\cdots,1,2,0,0,0,0,0)$ &
(10,11) + (1,55)   &
9(1,11)+9(16,1)  & $1 $ \\    
\cline{1-1}
$SO(10)\times SU(11) \times U(1)$ & + 2(1,1) & & \\
\hline\hline
$\frac13(1,\cdots , 1,0,0)$ &
(14,2,2) + (91,1,1)   &
9(1,1,1) + 9(14,2,1) & $1 $ \\      
\cline{1-1}
$SU(14)\times U(1) \times SU(2)\times SU(2)$ &  + 2(1,1,1) & + 
18(1,1,2)$^*$& \\ 
\hline
\end{tabular}
\end{center}
\caption{Perturbative $Z_2$ and $Z_3$, $Spin(32)/Z_2$, orbifold models .   The asterisk 
indicates twisted states involving left-handed oscillators. The last column 
shows the generic terminal gauge group $G_0$ after Higgsing.
\label{tabla2} }
\end{table}

\medskip

{\bf   ii) Smooth K3 Compactifications}

It is instructive to compare these orbifold vacua with the $D=6$, $N=1$ 
models obtained upon generic heterotic compactifications on 
smooth $K3$ surfaces in the presence of instanton backgrounds
\cite{kv, dmw, sw6d}.
In the $E_8\times E_8$ case there are instanton numbers $(k_1,k_2)$  
satisfying $k_1+k_2=24$, as required by anomaly cancellation.
It is convenient to define $k_1=12+n $, $ k_2= 12-n $
and assume $n\ge 0$ without loss of generality. For $n \leq 8$,
an $SU(2)$ background on each $E_8$ leads to $E_7\times E_7$ 
unbroken gauge group with hypermultiplet content 
\beq
\oh(8+n)({\bf 56},{\bf 1}) + \oh (8-n)({\bf 1}, {\bf 56})
+  62({\bf 1}, {\bf 1})
\label{e7e7}
\eeq
Due to the pseudoreal character of the ${\bf 56}$ of $E_7$,
odd values of $n$ can also be considered. For the models in the range 
$8< n \leq 11$, non-perturbative
small instanton considerations are needed (see 2.3). There is a last model 
for 
$n=12$, which is obtained by embedding all 24 instantons on one $E_8$. 
The reader may check how all the irreducible terms in the anomaly 
polynomial cancel and factorization of the residual  anomalies occurs.

Models with diverse groups can be obtained from these spectra by symmetry 
breaking. The group from the second $E_8$
does not possess, in general, enough charged matter to be completely broken.
Higgsing stops at some terminal group, depending on the value of $n$, 
with minimal or no charged matter \cite{kv, afiq}. 
For instance $E_8$, $E_7$, $E_6$, $SO(8)$, $SU(3)$ 
terminal groups are obtained for $n=12,8,6,4,3$ while complete breaking 
proceeds for $n=2,1,0$. On the other hand, the first $E_7$ can be completely 
Higgsed away. 
 For generic
gauge group $G= G_1\times G_2$ with $G_1$ and $G_2$ subgroups of the first and 
second $E_8$ obtained from backgrounds with instanton numbers $(12+n,12-n)$,
the following identity is satisfied 
\beq
\frac{{\tilde V}_{1}}{V_1}= \frac{n}{2}  \quad\quad ; \quad\quad   
\frac{{\tilde V }_{2}}{V_2}= -\frac{n}{2} 
\label{vtfor}
\eeq
These relations remain valid at each step of possible Higgsing.
{}From the anomaly polynomial it follows that the gauge kinetic 
terms are proportional to \cite{sagn}
\beq
-V_1 (e^{-\phi}+ \frac{n}2e^{\phi})\tr\, F_1^2 -
 V_2 (e^{-\phi}-\frac{n}2e^{\phi})\tr\, F_2^2 
\label{sagnot}
\eeq
where $F_i$ is the field strength of the unbroken group $G_i$ 
and $\phi$ is the scalar dilaton living in a $D=6$ tensor multiplet.
The coefficient of the gauge kinetic term for the second $E_8$ is
such that the gauge coupling diverges at \cite{dmw, sw6d} 
\beq
e^{-2\phi} = \frac{n}2
\label{phtr}
\eeq
This is a sign of a phase transition in which there appear tensionless
strings \cite{gh, sw6d, dlp}, as we will describe in section 3.5.

 In the last column of Table~\ref{tabla1} we show the instanton numbers
$(k_1,k_2)$ of compactifications yielding, upon Higgsing, 
a massless spectrum similar to the corresponding orbifold. We thus see that the
five $Z_3$ orbifolds of $E_8\times E_8$  are in the same
moduli space as generic $K3$ compactifications with $n=12,0,12,6,3$ 
respectively. The two $Z_2$ orbifolds correspond to $n=12,4$ respectively.
This connection between modular invariant orbifold models and
instanton backgrounds  is explained in more detail in ref.\cite{afiuv} .

In the $Spin(32)/Z_2$ case, embedding a total instanton number $k=24$ is required
to cancel gravitational anomalies. An $SU(2)$ background breaks the symmetry down 
to $SO(28)\times SU(2)$ with hypermultiplets in 
$10({\bf 28},{\bf 2})+65({\bf 1},{\bf 1})$.  Hence,
upon Higgsing, the generic group is $SO(8)$.  This class
of models is known to be  \cite{mv1, berkooz}
 in the same moduli space as $(k_1,k_2)=(16,8)$ 
compactifications of $E_8\times E_8$. As shown in Table~\ref{tabla2}, 
the first three  $Z_3$ orbifolds  of $Spin(32)/Z_2$  do have $SO(8)$ as 
generic group 
but the last two models have trivial gauge group after full Higgsing.
In fact,  it was already noticed in 
\cite{afiu} that the fourth $Spin(32)/Z_2$ $Z_3$ model could 
lead to complete Higgsing. Also,
in ref.~\cite{berkooz} the authors construct a heterotic $Z_2$ orbifold,
`without vector structure', in which the resulting $U(16)$ group can
be completely broken. In our language this $Z_2$ orbifold has embedding
$V=\frac14(1,\cdots,1,-3)$
 (third example in Table~\ref{tabla2}).
In general,  orbifold 
embeddings with vector structure have shifts $V$ such that
$MV=(n_1, \cdots , n_{16})$, whereas embeddings without vector structure 
have $MV=(n_1 + \oh, \cdots , n_{16} +\oh)$. Since $MV \in \Gamma_{16}$,
$\sum_I n_I = {\rm even}$ in both cases.

We have seen that the   $E_8\times E_8$ compactifications can be labeled by
a pair of instanton numbers $(k_1,k_2)$ with
$k_1=12+n$, $k_2=12-n$ and $n=0,\cdots ,12$.
Recently it has become clear that there are in fact different types of
$Spin(32)/Z_2$ instantons which are classified by the generalized second 
Stieffel-Whitney class \cite{berkooz}.
An analysis in terms of F-theory  \cite{aspfz2} has shown that in a 
general $Spin(32)/Z_2$ heterotic compactification, instantons with and without 
vector structure are present, their contribution to the total instanton 
number being respectively $8+4n$ and $16-4n$, with the integer $n$ 
satisfying $-2\leq n \leq 4$. 
A simple heterotic realization of this idea can be 
obtained by embedding a $U(1)\times SU(2)$ background in 
$SO(32) \supset  SU(16) \times U(1) \supset
SU(14)\times U(1)^{\prime} \times U(1) \times SU(2)$.
Then the $Spin(32)/Z_2$ vacua can be labeled by 
giving the pair of instanton numbers $(k_{NA}, k_A)$ 
with $k_{NA}=8+4n$ and $k_A=16-4n$.
The adjoint decomposition is
\beqa
{\bf 496} & = & 
({\bf 1},0,0,{\bf 3}) + ({\ov {\bf 14}},\oh,0,{\bf 2}) + 
({\bf 14},-\oh,0,{\bf 2}) + ({\bf 195},0,0,{\bf 1}) + 2({\bf 1},0,0,{\bf 1})
+ \nonumber  \\
& {} & ({\bf 1},1,\s22,{\bf 1}) + ({\bf 14},\oh,\s22,{\bf 2}) + 
({\bf 91},0,\s22,{\bf 1}) 
+  \nonumber \\
& {} & ({\bf 1},-1,-\s22,{\bf 1}) + ({\ov{\bf 14}},-\oh,-\s22,{\bf 2}) + 
({\ov{\bf 91}},0,-\s22,{\bf 1}) 
\label{auno}
\eeqa
where the two middle entries denote the $U(1)^{\prime}\times U(1)$ charges.
The massless spectrum that arises upon embedding $k_A=(16-4n)$ instantons in $U(1)$
and $k_{NA}=(8+4n)$ in $SU(2)$ is found
using the index theorem formulae \cite{gsw,afiu}. For
$-1\leq n\leq 2$ we find the following $SU(14)\times U(1)^{\prime} \times U(1)$ 
hypermultiplets
\beqa
& {} & (1-\frac n2)({\bf 1},1,\s22) + (1-\frac n2)({\bf 1},-1,-\s22) + 
(1-\frac n2) ({\bf 91},0,\s22) + \nonumber \\
& {} & (1-\frac n2)({\ov{\bf 91}},0,-\s22) + (6+n)({\ov{\bf 14}},-\oh,-\s22) + 
(6+n)({\bf 14},\oh,\s22) + \nonumber \\
& {} &(2+2n)({\bf 14},-\oh,0) + (2+2n)({\ov{\bf 14}},\oh,0) 
+ (33+8n)({\bf 1},0,0)
\label{ados}
\eeqa

For $n=3$ there are not enough instantons to support the $U(1)$ bundle. 
The corresponding instantons become small and give the spectrum of a 
pointlike instanton without vector structure \cite{aspfz2} (see section 
2.3). The resulting model has a gauge group $SO(28)\times SU(2)\times 
Sp(4)$, a hypermultiplet content
\beq
8({\bf 28}, {\bf 2},{\bf 1}) + 56({\bf 1},{\bf 1},{\bf 1}) + \oh({\bf 
28},{\bf 1},{\bf 8}) + ({\bf 1},{\bf 2},{\bf 8})
\label{atres}
\eeq
and one additional tensor multiplet.
For $n=4$,  instantons without vector structure 
 disappear and one just has the $SU(2)$ bundle 
with 24 instantons mentioned above. For $n=-2$, the situation is reversed, 
since there only remains a
$U(1)$ bundle with 24 instantons. The resulting  gauge group is $U(16)$ with
hypermultiplets
\beq
2({\bf 120}, \frac1{2\sqrt2}) + 2({\bf \ov{120}}, -\frac1{2\sqrt2})
+ 20({\bf 1},0)
\label{acuatro}
\eeq
For each value of $n$, appropriate sequential Higgsing produces chains of 
models that match similar $E_8\times E_8$ heterotic chains \cite{afiq}, 
for the same value of $n$, thus providing several identifications between 
compactifications of both heterotic strings. This equivalence is evident 
in the F-theory framework (see  chapter  4), since the Calabi-Yau spaces obtained upon 
Higgsing (taking generic polynomials in the fibration over $\IF_n$) are 
identical in both types of chains. By computation of ${\tilde V}/V$ it 
can be shown that the $Z_3$  models listed in Table~\ref{tabla2} correspond to 
$n=4,4,4,1,1$, respectively. The three $Z_2$ models correspond to 
$n=4,4,0$.

{\bf    iii)    T-Dualities Between $E_8\times E_8$ and
$Spin(32)/Z_2$,   $D=6$, $N=1$ Vacua  }

As discussed above, $E_8\times E_8$  and  $Spin(32)/Z_2$ 
compactifications corresponding to the same values of $|n|$ are in the
same moduli space. This means that they must be in some way $T$-dual.
This $T$-duality may be explicitly shown in some cases in  terms of
orbifold compactifications as we now discuss. $T$-duality in toroidal 
compactifications is already present in $D=9$ \cite{ginsparg}.
More concretely, one can show that the
$E_8\times E_8$ heterotic compactified in a circle of radius $R$ 
and in the  presence of a Wilson line  of the form $a={\frac 12} 
(1,1,1,1,0,0,0,0)(1,1,1,1,0,0,0,0)$,   is equivalent to the $Spin(32)/Z_2$ 
compactified in a circle 
with radius $1/R$ and Wilson line $a={\frac 
12}(1,1,1,1,0,0,0,0,1,1,1,1,0,0,0,0)$.
In both cases the gauge group is $SO(16)\times SO(16)$ and the 
spectrum and interactions are identical.

Things are not so immediate in $D=6$, $N=1$ theories since we have now 
a smaller number of supersymmetries and the gauge backgrounds (e.g., Wilson lines) 
are not arbitrary. However, some of the equivalences can still be easily
proven. The equivalences for the cases $n=4$ and $n=0$ 
were shown in terms of $Z_2$ orbifolds in ref.\cite{berkooz}.
We rephrase their discussion  in the language of the bosonic 
formulation of  the heterotic string.  

\medskip

{\it  T-duality of $n=4$ vacua}

Consider a $E_8\times E_8$  $Z_2$ orbifold with shift
$V={\frac 12}(1,1,1,1,0,0,0,0)(1,1,0,0,0,0,0,0)$.  This breaks the symmetry 
down to $SO(16)\times E_7\times SU(2)$.
As shown in Table \ref{tabla1}, 
this is an orbifold version of  a $(k_1,k_2)=(16,8)$  
$E_8\times E_8$  compactification (hence $n=4$).  
 Add now a  quantized Wilson line
$a={\frac 12}(0,0,1,1,1,1,0,0)(1,1,1,1,0,0,0,0)$ which is of the same type 
discussed above for the $D=9$ case.
  It is easy to check 
that these verify the modular invariance constraints. The unbroken gauge 
group is  $SO(8)\times SO(8)\times SO(12)\times SO(4)$. 
Consider now the $Z_2$, $Spin(32)/Z_2$ orbifold with shift
$V={\frac 12}(1,1,1,1,0,0,0,0,1,1,0,0,0,0,0,0)$. This breaks the symmetry 
down to $SO(12)\times SO(20)$.  As shown in Table \ref{tabla1}  upon full
Higgsing this model (like the $E_8\times E_8$ one) leads to a
generic $SO(8)$ group (hence $n=4$).
  Add now  the Wilson line
$a={\frac 12}(1,1,1,1,1,1,1,1,0,0,0,0,0,0,0,0)$ which is again of the same 
type as above. Again, this is a modular invariant model
with the same gauge group. It is easy to check that the massless and
massive spectra  of these two models are exactly the same. Thus the
$n=4$  $E_8\times E_8$ compactification and  the $Spin(32)/Z_2$ 
compactifications with vector structure are in fact $T$-dual.

\medskip

{\it T-duality of $n=0$ vacua}

A $n=0$ vacuum  in $E_8\times E_8$ is symmetric  in both groups, so
we have to construct now an orbifold model with this symmetry.
Consider first a  $E_8\times E_8$, $Z_2$ orbifold with shift 
$V={\frac 14}(-3,1,1,1,1,1,1,1)(1,1,1,1,1,1,1,1)$. It is easy to see that
this shift is equivalent in the $E_8\times E_8$ lattice to the
second one in   Table \ref{tabla1}, leading to an unbroken
$SO(16)\times E_7\times SU(2)$  gauge group. This is not symmetric in both 
$E_8$'s but  the model may be symmetrized 
\cite{afiq2} by adding a discrete 
Wilson line  of the same form as in the previous example, 
$a={\frac 12}(1,-1,-1,-1,0,0,0,0)(-1,-1,-1,1,0,0,0,0)$.
Now the gauge group is $U(8)\times U(8)$. There are hypermultiplets
in the untwisted sector transforming like 
$({\bf 28}+{\bf {\overline {28}}},{\bf 1})+({\bf 1},{\bf 28}+{\bf {\overline 
{28}}}) +4({\bf 1},{\bf 1})$ and  in 
the twisted sectors like  $8({\bf 8}+{\bf {\bar 8}},{\bf 1})+8({\bf 
1},{\bf 8}+{\bf {\bar 8}})$.  The 
reader may check that this is  indeed  a $n=0$ model  by recalling that 
$n=2 {\tilde V}/V$ and noting that 
${\tilde V}=2-2=0$ from eqs.(\ref{vtildes}).
Now, one can show that the same model may be constructed starting from the
$Spin(32)/Z_2$ orbifold  without vector structure  obtained from the
$Z_2$ shift  $V={\frac 14}(-3,1,1,1,1,1,1,1,1,1,1,1,1,1,1,1)$. If we add 
again the Wilson line $a={\frac 12}(1,1,1,1,0,0,0,0,1,1,1,1,0,0,0,0)$, the 
unbroken group
is again $U(8)\times U(8)$ and the hypermultiplet content is identical to the
previous $E_8\times E_8$ case.  This is not surprising since both models are
subject to the same Wilson line and the gauge group before the $Z_2$ 
twist is identical
in both cases ($SO(16)^2$).  In addition, the  shift $V$, when restricted
to this subgroup in both cases, is also identical.  So this shows that the
$n=0$ $E_8\times E_8$ compactifications are T-dual  to $Spin(32)/Z_2$
compactifications without vector structure.

These  $n=0$ vacua  are relevant  for
heterotic/heterotic duality in $D=6$, as we will discuss later on. 
In fact, $n=2$ vacua  present  also heterotic/heterotic duality.
It has been shown by using F-theory that both type or vacua are
equivalent. This would suggest that they are in some sense
$T$-dual. We will discuss this point further in chapter~4.

\subsection{Non-Perturbative  Heterotic Vacua and Small Instantons}

{\bf i) Small Instantons in $Spin(32)/Z_2$ Heterotic}

Consider first a standard $Spin(32)/Z_2$ heterotic compactification
on a {\it smooth K3}.  As indicated above, a consistent perturbative
background requires the presence of a total of 24 instantons.
However, when the size of one  of these instantons becomes small
something interesting happens \cite{witsm}.  The perturbative 
compactification with 
$k=23$ is anomalous. However, the heterotic dilaton diverges exactly at 
the location of the small instanton, no matter how small its asymptotic 
value is \cite{callan}, so one has to deal with strongly coupled 
dynamics. In this case, these dynamics yield a new non-perturbative
gauge symmetry $Sp(1)\simeq SU(2)$,along with hypermultiplets 
transforming  as ${\frac 12}({\bf 32},{\bf 2}) +({\bf 1},{\bf 1})$ with 
respect to the 
group $SO(32)\times Sp(1)$.  What happens is that at this point a Type-I
Dirichlet five-brane appears with precisely those world-volume fields 
\footnote{Actually the dynamical object is composed of {\em two} Type-I 
D-branes, with $SU(2)$ Chan-Paton factors.}. The world-volume fills the 
six uncompactified dimensions, so these fields appear in spacetime.
If we decompose the hypermultiplets with respect to the
unbroken subgroup of $SO(32)$, one can check that 
gauge and gravitational anomalies cancel. Thus
beyond perturbation theory the condition $k=24$ is replaced by 
\beq
k+n_B=24
\label{anombso} 
\eeq
where $n_B$ is now the number of dynamical $SO(32)$ five-branes, which can be
understood as small instantons \cite{witsm}. One such brane carries,
as we said,  an $Sp(1)$
vector multiplet, but when $r$ of them coincide at a point on the smooth
$K3$, the group is enhanced to $Sp(r)$. In general, the non-perturbative
group is $\prod Sp(r_i)$ with $\sum r_i = n_B$. The five-branes also carry
non-perturbative hypermultiplets. In particular, for each $Sp(r)$ there
appear 32 half hypermultiplets in the fundamental representation, together
with one hypermultiplet in the antisymmetric two-index representation
(decomposable as a singlet plus the rest). Cancellation of gauge anomalies
requires that the hypermultiplets in the fundamental representation to be 
also
charged under the perturbative gauge group that arises when $SO(32)$ is
broken by the background with instanton number $k=24-n_B$. 

This is just a particular example of a more general phenomenon. Further
possibilities appear if the small instanton sits not at a smooth K3 point but 
at an A-D-E orbifold singularity. In the case of the $SO(32)$ heterotic 
string, their dynamics corresponds to that of Type-I D-branes at 
singular points, which have been recently studied in \cite{intri,bi1,bi2}, 
based on previous results from ref. \cite{quivers}. Five-branes with a 
variety of world-volume field contents appear in this case.
Some examples \cite{afiuv} 
corresponding to $A_m$,  $m=1,\ldots ,5$  singularities are 
displayed in Table
\ref{tablex}.   Notice that in these cases, except for the case of 
$Z_2$ singularities without vector structure, the 
world-volume theories contain tensor multiplets, and not only
vector multiplets as in the smooth K3 case.  Thus small $Spin(32)/Z_2$ 
instantons on  singularities give transitions to Coulomb phases
parametrized by the real scalars in tensor multiplets.

\begin{table}[phtb]
\renewcommand{\arraystretch}{1.25}
\begin{center}
\footnotesize
\begin{tabular}{|c|c|c|c|}
\hline
$ \! Z_M \! $   &   Gauge Group   &   Hypermultiplets & $\! n_T \! $ \\
\hline
\hline
\multicolumn{4}{|c|}{Embeddings with vector structure 
} \\ 
\hline -- &  $Sp(\ell)$  &   ${\frac {32}2}(2\ell)+\ell(2\ell-1)$ & 0 \\
\hline
$Z_2$  &   $Sp(\ell)\times Sp(\ell+{\frac {w_1}2}-4) $  
 &  $w_0(2\ell,1)+w_1(1,2\ell+w_1-8)+(2\ell,2\ell+w_1-8)$ & 1 \\
\hline
$Z_2$ & $Sp(\ell)\times SO(2\ell+8) \quad [w_1=0]$ & $(2\ell, 2\ell+ 8)$ & 1 \\
\hline
$Z_3$  & $Sp(\ell)\times U(2\ell+w_1-8)$    
&    $w_0(2\ell,1)+w_1(1,2\ell+w_1-8)$ & 1 \\
 &   &  $ + (2\ell, 2\ell+w_1-8)+(1,(\ell+{\frac{w_1}2}-4)(2\ell+w_1-9))$ & \\
\hline
$Z_4$  &   $Sp(\ell)\times U(2\ell+w_1+w_2-8) $  &
  $w_0(2\ell,1,1)+ w_1(1, 2\ell+w_1+w_2-8,1)$ & 2 \\
              &    $\times  Sp(\ell+\frac{w_1}2+w_2-8)$  &
$+w_2(1,1,2\ell+w_1+2w_2-16)+(2\ell,2\ell+w_1+w_2-8, 1) $  & \\
             &       & $+(1, 2\ell+w_1+w_2-8,  2\ell+w_1+2w_2-16)$ & \\
\hline\hline
\multicolumn{4}{|c|}{Embeddings without vector structure} \\
\hline
$Z_2$  &   $U(2\ell)$  &  ${\frac {32}2}(2\ell)+2(\ell(2\ell-1))$ & 0 \\
\hline
$Z_4$  &   $U(2\ell)\times U(2\ell+u_2-8)$    &
   $u_1(2\ell,1)+u_2(1,2\ell+u_2-8)+ (2\ell, 2\ell+u_2-8)$ & 1 \\
    &   &   $ + (1,(\ell+{\frac{u_2}2}-4)(2\ell+u_2-9))+(\ell(2\ell-1),1)$ & \\
\hline
$Z_6$  &   $U(2\ell)\times U(2\ell+u_2+u_3-8)$    &
$u_1(2\ell,1,1)+u_2(1,2\ell+u_2+u_3-8,1)  + (\ell(2\ell-1),1,1)  $  & 2 \\
     & $\times U(2\ell+u_2+2u_3-16)$  &  
           $+u_3(1,1,2\ell+u_2+2u_3-16)+(2\ell,2\ell+u_2+u_3-8,1)$ & \\
   &   & $+ (1,2\ell+u_2+u_3-8,2\ell+u_2+2u_3-16)  $ & \\
      &     & $+(1,1,(\ell+{\frac{u_2}2}+u_3-8)(2\ell+u_2+2u_3-17))$ & \\
\hline
\end{tabular}
\end{center}
\caption{ Some world-volume theories of $SO(32)$ five-branes
 at $Z_M$ singularities   .   Here $w_\mu$ is the number of entries equal to $\frac{\mu}M$ in $V$ with
vector structure. Similarly, $u_\mu$ is the number of entries equal to 
$\frac{2\mu -1}{2M}$ in $V$ without vector structure.
\label{tablex}}
\end{table}

\smallskip

Recently non-perturbative $D=6$  heterotic orbifold models  have also  been
constructed  which  correspond to  the presence of small instantons 
either  moving on the bulk or stuck at the orbifold fixed points 
\cite{afiuv}. These are in some sense  toroidal orbifold versions of the
K3 vacua  with $k<24$ considered above. 
 We present an example
here which  contains five-branes stuck at $Z_2$ singularities.
 It is  a  $Z_2$  orbifold  of
heterotic $SO(32)$ which yields the same spectrum as the
$Z_2$ orientifold constructed by Dabholkar and Park \cite{dabol2}  
and  model C of Gopakumar and Mukhi \cite{gm}.
This  is a $D=6$, $N=1$
model with gauge group $SO(8)^8$, seventeen tensors and four hypermultiplets.
It can be obtained  in terms of F-theory compactified on the standard
$Z_2\times Z_2$ orbifold,  as a compactification of 
M-theory  on $T^5/Z_2\times Z_2$  and as a Type-IIB orientifold.
Here  we will obtain it  as a heterotic 
 $SO(32)$  $Z_2$ orbifold (with a non-modular
invariant  gauge shift  \cite{afiuv} ).  
 We will embed the $Z_2$ twist  in terms of a shift $V$ in 
the $\Gamma _{16} $ lattice  supplemented with two discrete Wilson lines
$a_1$ and $a_2$  as follows
\beqa
V\ =\quad  a_1 & =& {\frac 12} (1,1,1,1,1,1,1,1,0,0,0,0,0,0,0,0) \nonumber \\
a_2  &=&  {\frac 12} (0,0,0,0,1,1,1,1,1,1,1,1,0,0,0,0) 
\label{shigm}
\eeqa
The two Wilson lines break the symmetry down to $SO(8)^4$  whereas 
the $V$ shift projects out  all charged multiplets from the untwisted sector.
Only the four untwisted singlet hypermultiplets remain in that sector. The 
sixteen twisted sectors split into four sets of four fixed points each, which
are subject to shifts  $V$, $V+a_2$, $V+a_1+a_2$ and $V+a_1$  respectively.
The first three sets of  four fixed points are all similar, the corresponding 
shift has eight ${\frac 12}$ entries.  Thus $V^2=2$ and there are no massless
hypermultiplets at any of those twelve fixed points. Looking at
table~\ref{tablex},  we see that for $Z_2$ embeddings with vector structure 
one tensor  and a gauge group $Sp(\ell)\times Sp(\ell+\frac{w_1}2-4)$
appear \cite{intri}. Since in our case $w_1=8$, we get one tensor 
for each of the twelve fixed points and no enhanced gauge group for $\ell=0$.

The other four fixed  points with shift $V+a_1$ have a different behaviour.
Indeed, this shift is trivial and hence we have $w_1=0$ for those fixed points.
As remarked in ref.~\cite{bi2}, five-branes at a $Z_2$ singularity
with $w_1=0$  give transitions to a Coulomb phase with one tensor multiplet
and a gauge group $Sp(\ell)\times SO(2\ell+8)$
(see Table~\ref{tablex}). Thus, in our case, with $\ell=0$ 
at each of the fixed points we have altogether a non-perturbative
group $SO(8)^4$ and four tensor multiplets.
Putting  all the contributions together we get the total content
$SO(8)^8$, seventeen tensor multiplets and four singlet hypermultiplets.
Notice how the 16 twisted sectors are in a Coulomb phase, twelve of them
with $w_1=8$ yielding only tensors and the other four 
have $w_1=0$ yielding in addition the required non-perturbative $SO(8)^4$.

{\bf ii) Small Instantons in $E_8\times E_8$}

The physics of small instantons in $E_8\times E_8$ heterotic is
somewhat different  since the  strongly coupled limit of this theory 
is given by M-theory compactified on the segment $S^1/Z_2$.
Now the  M-theory five-branes play the role played by type I
Dirichlet  five-branes in the $SO(32)$ case. 
When compactifying the $E_8\times E_8$ heterotic on a smooth 
K3  with $n_B$  pointlike instantons, $n_B$  five-branes appear, with their
world-volume spanning the six uncompactified dimensions.
The world-volume theory includes one $D=6$ tensor 
multiplet and one singlet hypermultiplet. Altogether, each one
contains 5 real scalars which parametrize  the position of the
five-brane on  $K3\times S^1/Z_2$.
The five-branes are a source of torsion
so that in a case with $k_1$ instantons in the first $E_8$, $k_2$ 
in the second and $n_B$ five-branes at points in 
$K_3\times S^1/ Z_2$, the condition $k_1 + k_2 = 24$ is replaced by 
\beq
k_1+k_2+n_B=24
\label{ancant}
\eeq
For {\it smooth}  K3 compactifications, the physics when the instanton
becomes small  is very different  in $E_8\times E_8$ compared to the
$SO(32)$ case.  In the latter case the transition may be considered
as a standard Higgs phase in the sense that the process in which
the instanton recovers a finite size may be described as a Higgs effect 
in which the non-perturbative gauge symmetry $Sp(n_B)$
is  Higgsed away.  Also, the  $D=6$ theory living on the
five-brane  world-volume is infrared free.  In the $E_8\times E_8$ case
the small instanton theory is at a 
Coulomb phase parametrized by the real scalar in the tensor multiplet 
\cite{gh, sw6d}.  At the transition point, unlike the $SO(32)$ case,
there is a non-trivial scale invariant interacting $D=6$ field  theory  in 
which tensionless strings appear. Away from the transition point 
there is a massless tensor multiplet plus a hypermultiplet which, as we said
parametrize the position of the five-brane.

The above remarks seem to indicate that  there are 
 no enhanced non-perturbative gauge group  in
$E_8\times E_8$ compactifications.  However, this statement has to be
qualified in various ways.  First,   we know that $E_8\times E_8$ 
compactifications with $0\le n\le 4 $ are  T-dual  to  $Spin(32)/Z_2$ 
compactifications with the same $n$,   and we know that  in such
$Spin(32)/Z_2$ compactifications   {\it there are }  non-perturbative
enhanced gauge groups.  One thus concludes that  $E_8\times E_8$
K3 smooth compactifications  with $0\le n \le 4$  may have indeed 
non-perturbative enhanced gauge symmetries  in points of their moduli
space corresponding to small  instantons in the T-dual  $Spin(32)/Z_2$ 
model \cite{berkooz}.  The second qualification concerns the effect of $K3$
singularities. We already remarked how in the $Spin(32)$ case 
five-branes sitting at  singularities give rise  not only to additional
non-perturbative gauge groups but  also to tensors.  The opposite
can be said in $E_8\times E_8$: when five-branes sit at
A-D-E singularities not only tensors appear but also new non-perturbative
gauge groups.  This has been recently analyzed  in  \cite{am88}.

\section{Type-IIA $D=6,4$ Vacua and Type-II/Heterotic  Duality }

\subsection{ $D=4$, $N=2$ Type-II/Heterotic vacua}

In this and the following subsections we will make a detour from six 
dimensions to study $D=4$, $N=2$ heterotic and Type-IIA vacua, and the 
relations among them.

The supermultiplet structure of $D=4$, $N=2$ theories is
\beqa
Gravity & \rightarrow R(4)  = & \{ g_{\mu\nu}, \psi_\mu^{(+)}, 
\psi_\mu^{(-)}, 
B_\mu \} \quad\quad ; \quad\quad \mu, \nu = 0, \ldots, 3 \nonumber \\
Vector & \rightarrow V(4)  = & \{ A_\mu , \lambda^{(+)}, \lambda^{(-)} , 
2a \} \nonumber \\
Hypermultiplet & \rightarrow H(4)  = & \{ \chi^{(-)}, \chi^{(+)}, 4\phi 
\} ~. 
\label{multi4}
\eeqa

The presence of two kinds of scalars, those in vector multiplets and 
those in hypermultiplets, implies the existence of two branches in the 
moduli space. The Higgs branch is parametrized by the scalars in 
hypermultiplets, and is very similar to the six dimensional Higgs branch. 
Vevs for charged hypermultiplets trigger gauge symmetry breaking with 
reduction  (in general) 
of the rank. At a generic point in this phase, the gauge group 
is broken to a terminal one, which usually does not contain any charged 
matter. 
The Coulomb phase, on the other hand, is associated to the 
scalars in vector multiplets. Along this branch, gauge symmetry breaking of 
the non-abelian groups occurs, due to vevs associated to the scalars in the 
adjoint. Consequently, there is no rank reduction, 
and at a generic point in this moduli space the symmetry group reduces to
the Cartan subalgebra. Also, mass terms are generated for the charged 
hypermultiplets, due to their coupling to the scalars in vector 
multiplets, and they become heavy. 

Notice that $D=6$, $N=1$ theories reduce to $D=4$, $N=2$ ones upon 
compactification on a $T^2$. In this reduction, the multiplets 
(\ref{multi6}) decompose as 
\beqa
{\rm Gravity} & R(6) & \longrightarrow  R(4) + 2 V(4) \nonumber\\
{\rm Tensor} & T(6) & \longrightarrow  V(4) \nonumber \\
{\rm Vector} & V(6) & \longrightarrow  V(4) \\
{\rm Hyperm.} & H(6) & \longrightarrow  H(4) ~. \nonumber
\label{reducci}
\eeqa

Notice that the $D=6$ tensor multiplet goes over to a $D=4$ vector 
multiplet, so that the $D=6$ Coulomb branch we introduced in 
previous sections is naturally contained in the (larger) $D=4$ Coulomb 
branch.
 
These $D=4$, $N=2$ theories are 
non-chiral, and thus not subject to anomaly cancellation constraints. 
However, strong statements can still be made concerning their 
non-perturbative behaviour due to some non-renormalization theorems, 
analogous to the six dimensional ones, in the sense that the scalars in 
vector- and hypermultiplets are decoupled.

Heterotic $D=4$, $N=2$ vacua are obtained through compactification on 
$K3\times T^2$, along with the choice of a non-trivial gauge bundle over the 
internal space. In some cases this construction amounts simply to a 
compactification to six dimensions on $K3$ followed by a reduction on 
$T^2$ to $D=4$. The spectrum in this case is easily found by a 
decomposition of the $D=6$ spectrum (obtained as in section 2.2) 
following (\ref{reducci}). It is important to notice that in the $T^2$ 
compactification an additional $U(1)^4$ factor appears, from the 
graviphoton in the gravity multiplet, 
the tensor multiplet containing the dilaton, and the two vector multiplets 
associated to the torus moduli. Also, one gets an additional $U(1)$ for 
each  extra tensor multiplet in the six dimensional theory.
Upon decompactification of the torus, one recovers the initial $D=6$ model.

There are other models which cannot be understood this way, since their 
construction involves choosing a $T^2$ with its K\"ahler class frozen at 
a specific value \cite{kv}. The spectrum can be found through index 
theorems as well. Note that in this case, the decompactification of the 
torus is not possible, and the models are not related to $D=6$, $N=1$ 
constructions. In this sense, they are intrinsically four dimensional.

In both cases, the generic spectrum of the $D=4$, $N=2$
heterotic  theory on the 
Coulomb branch contains an abelian gauge symmetry $U(1)^{n_V+1}$ and 
$n_H$ neutral hypermultiplets.

Type-IIA $D=4$, $N=2$ vacua can be obtained by compactification on a 
Calabi-Yau threefold. The spectrum is easily found by Kaluza-Klein 
reduction of the $D=10$ Type-IIA supergravity. If $(h_{11}, h_{12})$ denote 
the Hodge numbers of the internal space, the gauge group is 
$U(1)^{h_{11}+1}$ 
(where the $h_{11}$ vector bosons are obtained by integration of the ten 
dimensional 3-form on the $h_{11}$ non-trivial 2-cycles, and the 
remaining vector is the graviphoton), and there are $h_{12}+1$ 
hypermultiplets (where $h_{12}$ come from integrating the 3-form over 
$h_{12}$ 3-cycles, and the remaining one contains the Type-IIA dilaton). 
These hypermultiplets are neutral with respect to the symmetry group.
Note that this construction seems to be purely four dimensional, so in 
principle one would not expect $D=6$, $N=1$ relatives of these models. We 
will see this is not the case if the CY space is elliptically fibered.

\subsection{ $D=4$, $N=2$ Type-II/Heterotic duality}

\label{dual4d}

At a generic point in the Coulomb branch, the spectrum of 
a heterotic $D=4$, $N=2$ compactification is very 
analogous to the kind of spectra found for Type-IIA compactifications on 
Calabi-Yau spaces. Actually a non-perturbative duality relation has been 
conjectured  \cite{fhsv, kv} 
to hold between both kinds of compactifications. A necessary 
condition for two models to be dual is the matching of their spectra 
\cite{kv}: 
\beqa
n_V & = & h_{11} \nonumber \\
n_H & = & h_{12}+1
\label{match}
\eeqa

It has also been determined \cite{klm, al} that the 
Calabi-Yau should be a $K3$ 
fibration, so that the base $\IP_1$ is a preferred (1,1)-cycle, which gives 
the mode 
dual to the heterotic dilaton. This fact allows for a very intuitive 
picture of the duality {\em via} the adiabatic argument outlined in 
\cite{vw}. Understanding the heterotic $K3\times T^2$ as a fibration of 
$T^4$ over $\IP_1$, and the Type-IIA Calabi-Yau as a fibration of $K3$ 
over $\IP_1$, the $D=4$ duality is obtained by fiberwise application of 
the six dimensional duality between the heterotic string on $T^4$ and the 
Type-IIA on $K3$. This $D=6$, $N=2$ equivalence is firmly established and 
better understood than the 
proposed $D=4$, $N=2$ ones, so one can hope that some features of the 
former persist in the latter, helping in understanding their richer dynamics.

For example, heterotic compactifications lead to enhanced non-abelian 
gauge symmetries whenever vevs for some scalars are set to zero. The 
Type-IIA mechanism required for generating such 
symmetries is inherited from an analogous phenomenon in the $D=6$, $N=2$ 
theory: a singularity develops on the internal manifold, so that Type-IIA 
2-branes wrapping the zero size 2-cycles give additional vector 
multiplets enhancing a product of $U(1)$ factors to a full non-abelian 
symmetry group. 

A strong check of the duality between the heterotic and Type-IIA 
theories is the construction of dual pairs verifying (\ref{match}) 
\cite{kv,fhsv,vw,aspair}. In some 
cases, the heterotic sides of several dual pairs appear to be connected 
by perturbative processes (typically the Higgs mechanism). Duality then 
requires that non-perturbative transitions must exist among the Type-IIA 
realizations of these models. Our aim in this chapter is to 
accumulate evidence in favour of this matching between 
the web of heterotic and Type-IIA vacua, and to learn about the 
non-perturbative dynamics it reveals.

An interesting subset of the moduli space can be explored by the 
construction of chains of models as in \cite{afiq}. The 
idea can be easily carried out in terms of the bundle construction of 
section 2.2. Consider the $D=4$, $N=2$ reduction of the $D=6$, $N=1$ 
spectrum (\ref{e7e7}). As already mentioned, we can Higgs the second 
$E_7$ down to a terminal group. We can then Higgs the first $E_7$ 
sequentially, lowering the rank of the gauge symmetry in units, and going 
to the Coulomb branch at each stage. This generates a chain of models 
whose spectra can be compared, {\em via} (\ref{match}), with the Hodge 
numbers of candidate Calabi-Yau duals.

Let us consider for example the $n=4$ case, for which the spectrum 
(\ref{e7e7}), once in $D=4$, $N=2$, is
\beqa
& {} & \hspace{1cm} E_7  \times E_7\times U(1)^4 \nonumber\\
& {} & 6({\bf 56},{\bf 1})  +  2({\bf 1}, {\bf 56}) + 62({\bf 1}, {\bf 1}) 
\eeqa
The second $E_7$ can be Higgsed down to $SO(8)$ with no matter. By going to 
the Coulomb 
phase, we get a model with a gauge symmetry $U(1)^{15}$ and 69 neutral 
hypermultiplets. The CY space reproducing this spectrum must have, from 
the matching conditions, $(h_{11},h_{12})=(14,68)$. Instead, we can go 
further 
along the Higgs branch, by the sequential breaking of the remaining $E_7$ 
to $E_6$, $SO(10)$, $SU(5)$, $SU(4)$, $SU(3)$, $SU(2)$ and to nothing. 
The Hodge numbers of candidate dual CY's are found to be $(13,79)$, 
$(12,88)$, $(11,95)$, $(10,122)$, $(9,153)$, $(8,194)$ and $(7,271)$, 
respectively. The 
easiest way of identifying such spaces is to look for these Hodge numbers 
in the lists of CY threefolds which are $K3$ fibrations. Actually, in the 
tables of \cite{klm} one finds candidate CY's for the last four 
elements in the chain, realized as the varieties
$\IP_5^{(1,1,4,8,10,12)}[20,16]$, $\IP_4^{(1,1,4,8,10)}[24]$, 
$\IP_4^{(1,1,4,8,14)}[28]$ and $\IP_4^{(1,1,4,12,18)}[36]$.

This exercise can be carried out for other values of $n$ \footnote{A 
subtlety 
arises for $n=9,10,11$, in which case there are not enough instantons in 
the second $E_8$ to support an $SU(2)$ bundle, so they are forced to be 
point-like.}. The Calabi-Yau varieties associated to the last steps of 
Higgsing for 
n=2,4,6,8,12 can be found in the tables of ref. \cite{klm}. 
Remarkably, a pattern in the weights defining these spaces is observed. 
The heterotic cascade breaking sequence

\beq
\cdots \rightarrow SU(4) \rightarrow SU(3) \rightarrow SU(2) \rightarrow
\emptyset
\label{casca}
\eeq

maps into the following sequence in the Type-II side
\beqa
 & \IP_5^{(1,1,w_1,w_2,w_3,w_4)}  \rightarrow
\IP_4^{(1,1,w_1,w_2,w_3)} \rightarrow
\nonumber \\
&  \IP_4^{(1,1,w_1,w_2,w_3+w_1)} \rightarrow
\IP_4^{(1,1,w_1,w_2+w_1,w_3+2w_1)}
\label{cypatt}
\eeqa

Moreover, these transitions can be recast in terms of $n$, as follows
\beqa
& {} & \IP_5^{(1,1,n, n+4,n+6, n+8)}[2n+12,2n+8]
\rightarrow  \IP_4^{(1,1,n,n+4,n+6)}[3n+12] \rightarrow
\nonumber\\
&{}& \hspace{0.5cm}
\IP_4^{(1,1,n,n+4,2n+6)}[4n+12] \rightarrow
\IP_4^{(1,1,n,2n+4,3n+6)}[6n+12]
\label{achain}
\eeqa

In the following section the construction of these CY's will be considered 
in a more appropriate framework, which allows to find CY spaces for any 
value of $n$ in the range $0\leq n \leq 12$, and moreover shows that this 
pattern remains valid for all $n$.

These regularities point towards 
the existence of transitions among the compactifications on these CY's. 
\cite{afiq, afiu} .
 They are generalizations of the conifold transition in 
\cite{conist,gms}. One starts from a CY corresponding to a model in 
the 
Coulomb branch. By setting scalars in vector multiplets to zero, some 
2-cycles on the CY collapse, leading to a singularity that produces a 
non-abelian enhancement of the symmetry. The singularity can then be 
smoothed by a deformation of the complex structure, i.e. by turning on 
vevs for scalars in hypermultiplets. This process is the dual picture of 
the sequential Higgsing we have introduced from the heterotic viewpoint.

There also exist processes connecting chains for different values of $n$ 
\cite{sw6d}. 
They consist, in the heterotic picture, on shrinking an $E_8$ instanton 
to zero size, and transforming it into a M-theory five-brane. Changing 
the vev of the scalar in its tensor multiplet the five-brane can be made to 
travel along the Coulomb branch until it is reabsorbed as an instanton on 
the other $E_8$. This process is purely six dimensional and moreover, 
cannot be described within the framework of field theory, since it is 
mediated by tensionless strings \cite{gh}. Note that the transformation 
of five-branes into finite size instantons realizes the first transition 
proposed in (\ref{transia}). The tensor present in the Coulomb phase 
turns into hypermultiplets (in adequate representations under the gauge 
group) in the Higgs phase.

The construction of chains can also be carried out from the heterotic 
$SO(32)$ bundle 
construction introduced in section 2.2. This time $n$ defines how the 
instanton number 24 splits between instantons with and without vector 
structure and has the range $-2\leq n\leq 4$. Starting from the unbroken 
group $SU(14)\times 
U(1)^{\prime}\times U(1)$ it is possible to Higgs the abelian factors, 
and start a sequential breaking of $SU(14)$. Using the spectra determined 
in section 2.2, it can be checked that for $-2\leq n \leq 2$ complete Higgsing
is possible, while for $n=4$ one ends up with 
a $SO(8)$ without matter. 
These coincide with the groups found in the $E_8\times E_8$ 
case for the same $n$. Actually, this coincidence is also obtained for 
the previous elements in the Higgsing chains, and can be understood as 
a consequence of the 
T-dualities between both heterotics, as advanced in section 2.2.
For $n=3$, on the other hand, the initial spectrum contains an 
additional tensor, associated to a small instanton. However, it can be 
argued from F-theory that coincidence with the $E_8\times E_8$ spectrum
is found once the tensor disappears non-perturbatively.

The mechanism we have described for changing the value of $n$ is also 
valid in the $SO(32)$ case. The essential point is the appearance of a 
tensor degree of freedom at some locus in the hypermultiplet moduli 
space, as has been mentioned in section 2.3. It has been checked from 
F-theory that one can travel along this Coulomb branch, and land on a 
different Higgs branch, where the tensor disappears and there has been 
a change of one unit in $n$ \cite{aspfz2}. The correct interpretation of 
this phenomenon in the $SO(32)$ heterotic is challenging, since there is 
no direct relation to M-theory. It certainly point towards a more unified 
picture of both heterotics.

\subsection{F-Theory-Heterotic Duality}

\label{dualfth}

In the discussion of the previous subsection the heterotic models we 
employed where essentially six dimensional, the dynamics did not depend 
on the $T^2$, while the Type-IIA construction is defined directly in four 
dimensions. The existence of a well defined decompactification limit 
imposes some condition on the geometry of the CY spaces, namely they must 
be elliptic fibrations. This has been determined from the Type-IIA 
viewpoint \cite{ag}, but we will motivate it from a different 
approach, F-theory \cite{fth,mv1,mv2}.

\subsubsection{Introduction to F-theory}  

A  new insight into several string dualities has 
been provided by
F-theory \cite{fth}, a construction that can be 
understood as a Type-IIB
compactification on a variety B in the presence of  
Dirichlet seven-branes. The
complex `coupling constant' $\tau=a+ie^{-\varphi/2}$, 
where $a$ is the RR 
scalar and $\varphi$ is the dilaton field, depends on 
space-time and is furthermore
allowed to undergo $SL(2,\IZ)$ monodromies around the 
seven-branes. This $\tau$
can be thought to describe the complex
structure parameter of a torus (of frozen K\"{a}hler class, since 
Type-IIB theory has no fields to account for it) varying over the
compactifying space $B$, and degenerating at the eight-dimensional 
submanifolds defined 
by the world-volumes of the seven-branes. The constraint of having vanishing 
first Chern class (the contribution of the seven-branes cancelling
that of the manifold $B$) forces the fibration of $T^2$ over $B$  
thus constructed to be an elliptic CY manifold $X$.
Thus, F-theory compactifications are defined only on elliptically fibered 
manifolds.

The compactification of F-theory on the product of such an elliptically 
fibered manifold $X$ and a circle $S^1$, lies on the
same moduli space as M-theory compactified on $X$ \cite{fth}. The 
result follows from the fact that
M-theory on $T^2$ is equivalent to the Type-IIB theory on $S^1$, the IIB 
coupling constant $\tau$ being equal to the modular parameter of the 
M-theory torus. By adiabatic fibering of this duality over the base B 
of the manifold X of interest, the desired result is obtained.
This equivalence has proved fruitful in encoding string dualities in lower 
dimensions, and, especially, in clarifying several phenomena in heterotic 
string compactifications.

\subsubsection{F-Theory/Heterotic duality} 

After compactification on an elliptic $K3$, F-theory gives a $D=8$ theory
conjectured to be dual to the heterotic string compactified on $T^2$
\cite{fth,senft, dieter}. 
The mapping of the moduli between both constructions is as follows. The 
size of the base $\IP_1$ is related to the heterotic dilaton whereas 
the 18
polynomial deformation complex parameters of the fibration match the 
heterotic
toroidal K\"{a}hler and complex structure moduli together with Wilson line
backgrounds. The K\"{a}hler class of the elliptic fiber on $K3$ has no 
physical meaning in  F-theory, and thus, no heterotic counterpart.

Fibering this model over another $\IP_1$ gives a family of F-theory
compactifications on CY three-folds which are $K3$ fibrations, with the $K3$
fibers admitting an elliptic fibration structure. The resulting base
spaces are the Hirzebruch surfaces $\IF_n$, which are fibrations
of $\IP_1$ over $\IP_1$, characterized by an integer $n$. These   
models are naturally conjectured to be dual to heterotic string
compactifications on $K3$ ($T^2$ fibered over $\IP_1$) with gauge bundles
embedded on $E_8 \times E_8$. For some values of $n$, it can also be   
related to $SO(32)$ heterotic string compactifications
\cite{berkooz, ag}, as we will mention at the end of the section. Upon 
toroidal compactification to 
$D=4$, $N=2$, heterotic/Type-IIA duality is recovered so that
F-theory provides an $N=1$, $D=6$ version of this duality. This implies
that the CY spaces dual to essentially six dimensional 
heterotic models should be elliptically fibered, so that they can be used 
for F-theory compactification.

The $D=6$, $N=1$ spectrum obtained from compactifying F-theory on a 
threefold can be partially determined using this relationship with 
Type-IIA compactifications. If we denote by $h_{11}(B)$ the number of 
(1,1)-forms of the base B, and $(h_{11}(X), h_{12}(X))$ the Hodge numbers of 
X, one can show \cite{fth} that the number of tensor multiplets is 
$h_{11}(B)-1$, the rank of the gauge group is $h_{11}(X)-h_{11}(B)-1$ and 
the number of neutral hypermultiplets is $h_{12}+1$.

Notice that in $D=6$ there is no vector Coulomb branch, so the non-abelian 
gauge groups do not break to their Cartan subalgebra. Further information 
about the gauge group is encoded in the curves of singularities of $X$ 
\cite{mv1,mv2}, i.e. in the overlapping D7-branes, in the IIB language. 
Charged hypermultiplets also remain in the massless spectrum, and they 
are usually associated to the intersection of the curves of 
singularities, i.e. correspond to open-strings stretching 
between different D7-branes.

The base space we are interested in, $\IF_n$,
being a  $\IP_1$ fibration over another $\IP_1$, has  
two K\"{a}hler forms, and thus the massless spectrum contains only one tensor
multiplet  (associated to the heterotic dilaton
\cite{mv1}). Consequently, except when
the singularities in the variety require a blow-up of the base for their
resolution, we will have $n_T=1$.

Our purpose is to find the F-theory duals of the
previously discussed heterotic models \cite{mv2}, by explicit construction 
of the fibrations as hypersurfaces in projective varieties.

\medskip

{\bf i) F-theory and heterotic $E_8\times E_8$}

The elliptic fiber can be realized as $\IP_2^{(1,2,3)}[6]$. Introducing
coordinates $z_1,w_1$ and $z_2,w_2$ for the two $\IP_1 $'s, $x,y$
for the torus, and two ${C}^*$ quotients  $\lambda, \mu$
to projectivize the affine spaces,
we obtain the following ambient space
\beq
\begin{tabular}{ccccccc}
& $z_1$ & $w_1$ & $z_2$ & $w_2$ & $x$ & $y$ \\
$\lambda : $ & 1 & 1 & 0 & 0 & 4 & 6  \\
$\mu$  : & $n$ & 0 & 1 & 1 & $2n+4$ & $3n+6$
\end{tabular}
\label{fiba}
\eeq
The hypersurface in this space is given by the fibration equation
\begin{equation}
y^2 = x^3 + f(z_1,w_1;z_2,w_2) x  + g(z_1,w_1;z_2,w_2)
\label{eq:fibr1}
\end{equation}
where $f$ and $g$ are polynomials such that the equation
is invariant under the ${C}^*$ actions, and the 
variety is well defined on the projective space.
Notice that for  given $\omega _i, z_i$ this equation describes  a torus and
the complete equation is thus an elliptic fibration.
It can be shown that for $n>12$ the variety described by (\ref{fiba}) and 
(\ref{eq:fibr1}) does not fulfill the CY condition (in particular, the
associated Newton polyhedron ceases to be reflexive), so that there are
13 possible spaces.

For a fixed value of $n$ the moduli space of the fibrations (\ref{eq:fibr1}), 
parametrized by the coefficients of the polynomials $f$, $g$, reproduces the 
Higgs branch of the heterotic $E_8\times E_8$ 
model corresponding to embedding $(12+n,12-n)$ instantons in $E_8\times 
E_8$. 
Coulomb branches associated to tensors are obtained by blowing up the 
base $\IF_n$. 

The $E_8\times E_8$ structure is manifest for a particular choice of 
polynomials \cite{mv2}, leading to
\begin{center}
\beqa
& {} & \hspace{1cm} y^2  =  x^3 + f_8(z_2,w_2) z_1^4 w_1^4 \, x \nonumber\\ 
& {} & + g_{12-n}(z_2,w_2)  z_1^7 w_1^5  +  g_{12}(z_2,w_2) z_1^6 
w_1^6 + g_{12+n}(z_2,w_2) z_1^5 w_1^7 
\label{e8e8}
\eeqa
\end{center}
This model has gauge group $E_8\times E_8$,
since $z_1=0$ and $z_1=\infty$ ($w_1=0$) are two curves of $E_8$ 
singularities. The number of independent 
parameters can be computed to be 44. Also 24 blow-ups on the base space 
are required, so there are 24 tensor degrees of freedom. This model 
corresponds to 
the heterotic $E_8\times E_8$ with 24 pointlike instantons.

From this very special point in moduli space, one can reach more generic 
ones by allowing for more generic polynomials $f$, $g$. This gives finite 
size to the small instantons. Even though the 
intermediate steps are also interesting e.g. to understand how charged 
matter 
is encoded in the equations \cite{cf, bikmsv} we will mainly center on 
the models obtained upon maximal Higgsing, i.e. on the most generic 
polynomials.

To this end we expand the polynomials
$f,g$ in powers of $z_1,w_1$
\beqa
f(z_1,w_1;z_2,w_2) & = &
\sum_{k=-4}^{4} z_1^{4+k} w_1^{4-k} f_{8-nk}(z_2,w_2)  \nonumber \\
g(z_1,w_1;z_2,w_2) & = &
\sum_{l=-6}^{6} z_1^{6+l} w_1^{6-l} g_{12-nl}(z_2,w_2)
\label{expan1}
\eeqa
where subscripts
denote the degree of the polynomial in $z_2$, $w_2$ (only non-negative
degrees are admitted).

For $n \neq 0, 1$, we can dehomogeneize with respect to $w_1$
using one of the ${C}^*$ quotients, and the variety can be represented by
the hypersurface $\IP_4^{(1,1,n,2n+4,3n+6)}[6n+12]$. These coincide
with the last elements of the chains of section 3.2, showing they are 
elliptically fibered, as required. Furthermore, for {\em all} values of $n$, 
the Hodge numbers of the fibration do match the
matter spectrum of heterotic models on $K3\times T^2$ with SU(2) bundles of
$(12+n, 12-n)$
instanton number embedded in $E_8\times E_8$, upon maximal Higgsing and
moving to the Coulomb phase \cite{afiq, cf}. Thus, one identifies Type-IIA
compactifications on these spaces as duals of the heterotic
constructions in $D=4$, or equivalently the F-theory
compactifications as duals of the heterotic models in $D=6$ 
(decompactifying the $T^2$).

Let us illustrate with an example which further checks can be performed 
to confirm that this construction provides the required CY's. Take the 
$n=4$ case in eqs.~(\ref{expan1}). The gauge group comes from the generic 
singularity type along $w_1=0$. Locally (i.e. to lowest order in $w_1$) 
the fibration can be written
\beq
y^2 = x^3 + w_1^2 f_0 x + w_1^3 g_0
\eeq
which is a $D_4$ singularity
(the last term is an irrelevant deformation),  giving rise to  a $SO(8)$ 
gauge symmetry. 
Moreover, there are no additional D7-branes intersecting the curve 
$w_1=0$  ($f_0$ is a constant) 
and consequently no charged matter. Also, the total number of 
neutral 
hypermultiplets is obtained from $h_{12}$, and matches the heterotic 
result. A stronger check consists on counting independently the moduli 
associated to 
each initial $E_8$, and the $K3$ moduli \cite{bikmsv}. This 
computation yields a number in agreement with the index theorem for 
instantons embedded in $E_8$ used in 
the heterotic construction. In particular, let us compute the number of 
independent monomial deformations for $k,l<0$ in (\ref{expan1}): there 
are 76 monomials coming from $f_{12}$, $f_{16}$, $f_{20}$, and $f_{24}$, 
and 162 from $g_{16}$, $g_{20}$, $g_{24}$, $g_{28}$, $g_{32}$, and 
$g_{36}$. One must, however, substract 5 which can be eliminated by 
redefinitions $z_1 \to z_1 + P_{4}(z_2,w_2)$, and a last one from a 
global scaling of the polynomial equation. The result, 232, is equal to 
$30(12+4)-248$, the number of moduli for 16 $E_8$ instantons. Similar 
exercises show the mentioned agreement in all cases. Thus, every 
detail in the $D=6$ spectrum can be understood in the F-theory framework.
It has also been shown that by restriction of the polynomial coefficients in 
(\ref{expan1}) one can 
reproduce the CY spaces of the remaining elements in the chains, with 
results in complete agreement with heterotic expectations \cite{cf,bikmsv}. 

The F-theory description is also fruitful in providing a 
detailed description of the $D=6$ transitions mediated by tensionless 
strings (see section 3.5). The simplest example which we have already 
encountered is 
the transition changing the value of $n$. In the F-theory context this 
amounts to a transition $\IF_n \to \IF_{n\pm 1}$, which is realized by 
blowing up the base at a point and then blowing down a curve of 
self-intersection $(-1)$ \cite{sw6d,mv2}. In the intermediate step a new 
K\"ahler class is 
introduced on the base, associated to the additional tensor from the 
M-theory five-brane. Also, at the boundaries of the Coulomb branch, a 
Type-IIB 3-brane wraps around the collapsed (1,1) cycles generating the 
tensionless string in $D=6$.

\medskip

{\bf F-theory and heterotic $Spin(32)/Z_2$}

The heterotic $Spin(32)/Z_2$ admits a F-theory description as well, by means 
of elliptic fibrations with two sections, in 
order to have $Spin(32)/Z_2$ (instead of $SO(32)$) as the gauge group 
\cite{ag,aspfz2}. 
These can be obtained by particularizing the polynomials $f$, $g$ in 
(\ref{eq:fibr1}) so that the fibration takes the factorized form
\beq
y^2 = (x-p(z_1,w_1;z_2,w_2)) (x^2 + p(z_1,w_1;z_2,w_2)\; x + 
q(z_1,w_1;z_2,w_2)) 
\label{fibfact}
\eeq
The heterotic model with unbroken $SO(32)$ is reproduced for the 
particular choice of polynomials \cite{ag}
\beqa
p(z_1,w_1;z_2,w_2) & = & B_{4+2n} (z_2,w_2) w_1^4   \nonumber\\
q(z_1,w_1;z_2,w_2) & = & A_{8+4n} (z_2,w_2) w_1^8  - 4 B_{4+2n} (z_2,w_2) 
C_{4-n} (z_2,w_2) w_1^5 z_1^3 \nonumber\\ 
& & -  2 C_{4-n}(z_2,w_2)^2 w_1^2 z_1^6 
\label{so32}
\eeqa
It can be checked that $w_1=0$ is a curve of $D_{16}$ singularities, but 
making this evident requires a non-trivial change of variables. 
There also appear several $Sp(k_i)$ factors, 
with $\sum k_i=8+4n$, with matter content coinciding with that of 
$SO(32)$ small instantons at smooth points (see table~\ref{tablex}). Also 
$4-n$ blow-ups of the base are required at $w_1=0$ and the zeroes of 
$C_{4-n}(z_2,w_2)$. This leads to $4-n$ tensors associated to the K\"ahler 
classes 
of the new curves, and a copy of $Sp(4)$ (with some matter content) each, 
since the fibers over the exceptional divisors are singular. This 
contribution is generated from $4-n$ groups of four small instantons 
each, at 
$Z_2$ singular points in the heterotic $K3$ (see table~\ref{tablex}). 
The number of neutral hypermultiplets comes up to be consistent with 
gravitational anomaly cancellation. This family of models illustrates 
how the different heterotic $Spin(32)/Z_2$ compactifications, with 
splitting of instantons defined by $n$, arise in F-theory.

As in the $E_8\times E_8$ case, one can give the instantons a finite 
size by allowing for more generic 
polynomials in the fibration. These deformation fall in two classes, 
those which respect the factorized form~(\ref{fibfact}) and those which 
do not. If one is to remain within theories with a $SO(32)$ heterotic 
interpretation, one must be careful in turning on the latter, as we will 
show below. Let us for the time being expand the most generic polynomials 
of the first kind
\beqa
p(z_1,w_1;z_2,w_2) & = & \sum_{k=-2}^2 p_{4-kn}(z_2,w_2) z_1^{2+k} 
w_1^{2-k} \nonumber\\ 
q(z_1,w_1;z_2,w_2) & = & \sum_{k=-4}^4 q_{8-kn}(z_2,w_2) z_1^{4+k} 
w_1^{4-k} \\ 
\label{expan32}
\eeqa
stressing again they will {\em not} allow us to explore the whole of the 
F-theory moduli space.
The models obtained for the most generic polynomials are closely similar 
to those in the $E_8\times E_8$ case (actually, when one allows for 
non-factorizable deformations the spaces are identical). This fact was 
already noticed in the heterotic construction of section 2.2, and is 
behind the many T-dualities between compactifications of both heterotics 
on $K3$. For $-2\leq n\leq 2$ Higgsing can proceed completely, while for 
$n=4$ a terminal $SO(8)$ without matter is found. The $n=3$ case is 
interesting, its local description near $w_1=0$ being
\beq
y^2 = x^3 + (q_2(z_2,w_2) - p_1(z_2,w_2)^2) w_1^2  \; x - p_1(z_2,w_2) 
q_2(z_2,w_2) w_1^3
\eeq
which corresponds to a semi-split $D_4$ singularity \cite{bikmsv}, 
leading to $SO(7)$ with two spinorials. This does not coincide 
directly with the $SU(3)$ without matter found in the $E_8\times E_8$ 
case, but matches it upon Higgsing.

As in the $E_8\times E_8$ case, all kind of transitions within a given 
chain, or between chains which differ in the value of $n$ are possible, 
by application of the same geometrical operations in the corresponding 
CY's. As remarked at the end of section 4.2, this points towards  
the idea that both 
heterotic construction lead to identical models, and only differ in the 
interpretation of the moduli.

\subsection{The A,B,C,D chains}

The construction of the $SU(2)$ bundle backgrounds in $E_8\times E_8$ 
and $SO(32)$ heterotics has been a useful tool to understand a large 
class of 
vacua in the moduli space of $D=6,4$ string vacua. We would like to 
stress the importance of the initial observation of the regularities in 
the weights of the CY's dual to some of these models.

This is enough motivation to search for similar patterns in the 
tables of \cite{klm}. These were already noticed in \cite{afiq} and 
further analyzed both from the F-theory and heterotic points of view in 
\cite{afiu}. The regularities found are shown in table~\ref{abcdpatt}, 
already recast in terms of an integer $n$. The family A corresponds to 
the chains studied in the previous section. Our purpose in the following 
is to understand the structure of these CY spaces, and to find their 
heterotic duals.

\begin{table}[htb]
\renewcommand{\arraystretch}{1.25}
\begin{center}
\begin{tabular}{|c|l|l|}
\hline
$r$ & \hspace{1.4cm}A  &
\hspace{1.4cm}B
\\[0.2ex] \hline
4 & $\IP_5^{(1,1,n, n+4,n+6, n+8)}$ & {} \\[0.2ex]
3 & $\IP_4^{(1,1,n,n+4,n+6)}$ &
$\IP_5^{(1,1,n,n+2,n+4,n+6)}$ \\[0.2ex]
2 & $\IP_4^{(1,1,n,n+4,2n+6)}$ &
$\IP_4^{(1,1,n,n+2,n+4)}$ \\[0.2ex]
1 & $\IP_4^{(1,1,n,2n+4,3n+6)}$ &
$\IP_4^{(1,1,n,n+2,2n+4)}$  \\[0.2ex]
\hline
\hline
$r$ &\hspace{1.4cm}  C  &
\hspace{1.4cm} D
\\[0.2ex] \hline
2 & $\IP_5^{(1,1,n,n+2,n+2, n+4)}$ & {} \\[0.2ex]
1 & $\IP_4^{(1,1,n,n+2,n+2)}$ &
$\IP_5^{(1,1,n,n+2,n+2, n+2)}$ \\[0.2ex]
\hline
\end{tabular}
\end{center}
\caption{ Structure of the A,B,C and D chains.}
\label{abcdpatt}  
\end{table}

We first note that these CY's are $K3$ fibrations, with the $K3$'s being 
elliptic fibrations. The
$T^2$ fibers are realized as $\IP_3^{(1,2,3)}[6]$, $\IP_3^{(1,1,2)}[4]$, 
$\IP_3^{(1,1,1)}[3]$ and $\IP_4^{(1,1,1,1)}[2,2]$ for the A, B, C and D 
families, respectively. This means that the models can be made sense of 
in $D=6$ through compactification of F-theory on them.

The elliptic fibration structure can be analyzed following the steps 
studied for the A models. For example, the CY's of the B family are 
defined in the ambient space given by
\beq
\begin{tabular}{ccccccc}
& $z_1$ & $w_1$ & $z_2$ & $w_2$ & $x$ & $y$ \\
$\lambda$ & 1 & 1 & 0 & 0 & 2 & 4 \\
$\mu$ & $n$ & 0 & 1 & 1 & $n+2$ & $2n+4$
\end{tabular}
\label{fibb}  
\eeq
and the hypersurface is provided by the equation
\begin{equation}
y^2 = x^4 + f(z_1,w_1;z_2,w_2) \, x^2  + g(z_1,w_1;z_2,w_2) \,x +
h(z_1,w_1;z_2,w_2)
\label{eq:fibr2}
\end{equation}

In each case $n$ is restricted by the condition that the set of weights
lead to a well defined CY space. For type A, $n \leq 12$ in agreement
with the heterotic construction. For types B and C, the weights
correspond to reflexive polyhedra only for $n \leq 8$ and $n \leq 6$
respectively. For models D, $n \leq 4$ is expected.

Let us consider some general features in their spectra.  In 
table~\ref{thodge} (from ref.\cite{afiu}) we 
show the Hodge numbers of the last elements in the chains. 
The bases of the fibrations are in most cases Hirzebruch surfaces 
$\IF_n$, but in the cases signalled with an asterisk some 
blow-ups of the base are required. So in general we get just one tensor 
multiplet. This fact and the existence of the discrete parameter $n$ 
suggest there may exist a heterotic construction in terms of bundles 
embedded in $E_8\times E_8$, with $n$ defining how the instanton number 
splits.

\begin{table}[htb]
\small
\renewcommand{\arraystretch}{1.25}
\begin{center}
\begin{tabular}{|c|cccc|}
\hline
$n$ & A & B & C & D\\[0.2ex]
\hline
\hline
0 & (243,3) & (148,4) & (101,5) & (70,6) \\
1 & (243,3) & (148,4) & (101,5) & (70,6) \\
2 & (243,3) & (148,4) & (101,5) & (70,6) \\
3 & (251,5) & (152,6) & (103,7) & $(70,10)^*$ \\
4 & (271,7) & (164,8) & (111,9) & (76,10) \\
5 & (295,7) & (178,10) & $(120,12)^*$ & {} \\
6 & (321,9) & (194,10) & (131,11) & {} \\
7 & (348,10) & $(210,12)^*$ & {} & {} \\
8 & (376,10) & (227,11) & {} & {} \\
9 & $(404,14)^*$ & {} & {} & {} \\
10 & $(433,13)^*$ & {} & {} & {} \\
11 & $(462,12)^*$ & {} & {} & {} \\
12 & (491,11) & {} & {} & {} \\
\hline
\end{tabular}
\end{center}
\caption{ Hodge numbers $(b_{21}^1, b_{11}^1)$ for the terminal spaces.}
\label{thodge}
\end{table}

We also see there is a general feature in the Hodge numbers in 
table~\ref{thodge}. For a given $n$, as one follows the  
sequence $A \to B \to C \to D$ (when possible) $h_{11}$ increases 
typically in one unit, while $h_{12}$ decreases in diverse amounts (which 
also follow certain numerological patterns, which we omit for the sake of 
brevity). 
The natural explanation is that the new families have an enhanced gauge 
symmetry: this increases the rank of the gauge group (related to 
$h_{11}$) and lowers the number of neutral hypermultiplets (related to 
$h_{12}$), since some previously neutral hypermultiplets can become charged 
with respect to the new symmetry.

The last piece of information comes from a detailed analysis of the 
fibration equations. Following the analysis performed for the 
family A of models, one can obtain the terminal gauge groups upon 
maximal Higgsing. The gauge 
group singularities exist only for $n \geq 2$ and are located at the 
curve $w_1=0$, just like in the case of the family A. The actual groups 
usually coincide with the terminal groups for chains A for the same $n$. 
One can also count the number of moduli coming from each 
(conjectured) bundle in $E_8$. This computation reveals the nature 
of the underlying groups before breaking through instantons takes place. 
For example, the chains B yield [18(9+n)-133] moduli for the gauge factor 
which 
is completely broken, pointing out the existence of an initial $E_7$ gauge 
symmetry.  Indeed, the dimension and Coxeter number of $E_7$ are
133 and 18 respectively.
Similarly, $E_6$ and $SO(10)$ structures underlay C and D 
models, respectively.
Observe that there seems to be some missing instanton number, also 
reflected in the smaller range of possible $n$'s, 
suggesting it must somehow be associated to the generation of the enhanced 
symmetry. Just to finish, concerning the nature of this extra groups, we 
are forced to accept they are not non-abelian, otherwise would have been 
detected as curves of singularities on the CY spaces. Abelian factors, 
however, appear in a much more elusive way, related to the rank of the 
Mordell-Weyl group of the elliptic fibration \cite{mv2}. 

This information serves as a guide for the construction of the heterotic 
duals. We will show how the desired new $D=6$ heterotic models
can be most readily obtained by considering generic
$H\times U(1)^{8-d}$  backgrounds in each $E_8$, with $H$ some
non-Abelian factor \cite{afiu}.

For example, embedding $SU(2)\times U(1)$ backgrounds
with instanton numbers $(k_1,m_1;k_2,m_2)$ in both $E_8$'s gives the
following $E_6\times U(1) \times E_6\times U(1)$ spectrum
\beqa
& {} &\big \{ 
\ds{\frac16(3k_1 + m_1 -12)} \ds{({\bf 27},\frac1{2\sqrt3};{\bf 1},0) +}
\nonumber\\[2mm]   
& {} &
 \ds{\frac16(3k_2 + m_2 -12)} \ds{({\bf 1},0 ; {\bf 27},\frac1{2\sqrt3})+}
\nonumber \\[2mm]
& {} &
\ds{\frac13(m_1 -3)} \ds{({\bf 27},-\frac1{\sqrt3};{\bf 1},0) +}
\nonumber\\[2mm]
& {} &
\ds{\frac13(m_2 -3)} \ds{ ({\bf 1},0 ; {\bf 27},-\frac1{\sqrt3}) + }
\nonumber\\[2mm]
& {} &
\ds{\frac12(k_1 + 3m_1 -4)}\ds{({\bf 1},\frac{\sqrt3}{2};{\bf 1},0)} +
\nonumber\\[2mm]
& {} &
 \ds{\frac12(k_2 + 3m_2 -4)} \ds{({\bf 1},0 ; {\bf 1},\frac{\sqrt3}{2}) +}
{\rm c.c.} \big \} +
\nonumber\\[2mm]
& {} &
\ds { ( (2k_1-3)+ (2k_2-3)}
\ds { + 20 )({\bf 1},0 ; {\bf 1},0)}
\label{espuna}
\eeqa
In this case gravitational anomalies cancel as long as
$k_1+m_1+k_2+m_2=24$.

At this point a brief comment concerning $U(1)$ anomalies is in order.
It is easy to check that $U(1)$'s in this class of theories are in
general anomalous. More precisely, one finds that the anomaly 8-form 
$I_8$  does not generically factorize into a product of two 4-forms,
so that the Green-Schwarz mechanism  cannot  cancel  the residual anomaly.
Instead one finds that the linear combination of $U(1)$ charges
\beq
 Q_f  = \cos\theta \, Q_1  + \sin\theta \, Q_2
\label{lacomb}
\eeq
leads to a factorized $I_8$  as long as
\beq
\sin^2\theta  = {{m_2}\over {m_1+m_2}} \quad\quad ; \quad\quad
\cos^2\theta  = {{m_1}\over {m_1+m_2}}
\label{normal}
\eeq
independently of the values of $k_{1,2}$.
Thus, for given $m_{1,2}$, there is a linear combination of both $U(1)$'s
which is anomaly-free  but the orthogonal combination is not.
Thus, somehow, the latter combination must be spontaneously broken.
Indeed, a mechanism by which this can take place was suggested in
refs.~\cite{witso,gsw} for analogous compactifications. The idea is
that in $D=10$ the kinetic term of the $B_{MN}$ field contains a piece
\beq
H^2 \simeq  (\, \partial _{\mu }B_{ij}\ +\ A^1_{\mu }
\langle F^1_{ij}\rangle \ + \  \ A^2_{\mu }\langle F^2_{ij}\rangle \, )^2
\label{truco}
\eeq
where the indices  $i,j$ live in the four compact dimensions. Notice that one
linear combination of $A^1_{\mu }$ and $A^2_{\mu }$  will become massive
by swallowing a $B_{ij}$ zero mode.

From eq.(\ref{espuna}) one 
notes  that  in the presence of the $SU(2)$ bundles the values of $m_{1,2}$
are forced to be multiples of $3$ in order to have half-integer numbers of
$({\bf 27} + \ov{{\bf 27}})$ and also $m_{1,2}\geq 3$.
Thus the simplest class of models of this type will have instanton numbers
$(k_1,3;k_2,3)$  and the unbroken $U(1)_f$ is in this case the diagonal
combination $U(1)_D$.
The fact that $k_1+k_2=18$, instead of $k_1+k_2=24$
(as in the case without $U(1)$
backgrounds), hints at the required heterotic duals
of models of type B. Indeed, in these models, the
range for the values of $n$ is smaller ($n\leq 8$) and this is
probably the case here since the range for $k_{1,2}$ is also smaller.
Moreover, models B have a number of vector
multiplets one unit higher compared to the corresponding chain A
elements. This is precisely the case here, due to
the presence of the extra $U(1)_D$.
These arguments are compelling enough to consider this sort of
heterotic constructions in more detail. Let us sketch how
upon sequential Higgsing of the non-Abelian symmetries the spectrum
in (\ref{espuna}) one does in fact reproduce chains of type B.
In analogy with the usual situation, we will label the
models in terms of the integer
\beq
n  =  k_1\ +\ m_1 \ - \ 12
\label{nuevon}
\eeq
where we assume without loss of generality that  $k_1+m_1\geq 12$.  We
choose $m_1=m_2=3$ as before so that $k_1+k_2= 18$.
 We now set up the derivation of the spectrum implied
by (\ref{espuna})
upon maximal Higgsing of non-Abelian symmetries. The results of course
depend on $n$ or equivalently on the pair $(k_1,k_2)$. The strategy is
to first implement breaking of the second $E_6$ together with $U(1)_D$
to $G_0 \times U(1)_X$, where $U(1)_X$ is
the appropriate `skew' combination of $U(1)_D$
and an $E_6$ Cartan generator. Since $k_1 \geq 9$, the first $E_6$ together
with $U(1)_X$ can then be broken to another `skew' $U(1)_Y$. The terminal
gauge group is therefore $G_0 \times U(1)_Y$ which by construction has
a factorized anomaly polynomial. Except for $n=5$, the terminal matter
consists of $G_0$ singlets charged under $U(1)_Y$ plus a number of
completely neutral hypermultiplets. The final step is to perform a
toroidal compactification on $T^2$ followed by transition to the
Coulomb phase. This allows us to compare the resulting spectrum of
vector and hypermultiplets with the Hodge numbers of candidate dual.
The agreement between the spectra found for the different
values of $n$ and the corresponding Type-II compactifications
is perfect. We refer the reader to \cite{afiu}  for further details.  

One can also
 consider $SU(2)\times U(1)^2$ backgrounds in each $E_8$. The $U(1)$'s
are embedded according to the branchings $E_8\supset SO(10)\times SU(4)$  and
$SU(4)\supset SU(2)\times SU(2)_A \times U(1)_B$
$\supset SU(2)\times U(1)_A\times U(1)_B$.
The distribution of instanton numbers is chosen to be
$(k_1,m_{1A},m_{1B};k_2,m_{2A},m_{2B})=(k_1,3,2;k_2,3,2)$, which can be shown
to guarantee a consistent spectrum. Notice
that anomaly cancellation requires $k_1+k_2=14$  (in the absence of extra
tensor multiplets from small instantons). The unbroken gauge group at
the starting level is $SO(10)\times U(1)^2\times SO(10)\times U(1)^2$.
In this case the diagonal combinations $Q_{AD} =  (Q_{1A} + Q_{2A})$
and $Q_{BD} =   (Q_{1B} + Q_{2B})$ are anomaly-free
whereas their orthogonal combinations are anomalous and are expected
to be Higgsed away by a mechanism analogous to that explained before.
Considering the different allowed values of $k_i$ and $m_{iA},m_{iB}$
one again finds perfect agreement with the spectra found
for the type C chains of Calabi-Yau compactifications.
Equally satisfactory results are found for the type D chains.

To end this section, let us stress that the new elliptic fibrations (or the 
heterotic non-semisimple backgrounds) allow us to explore a new direction 
in moduli space, that of enhancing of abelian gauge groups. In the 
heterotic description it is evident that all the different families are 
connected through Higgsing of the $U(1)$ factors. It has been shown in 
\cite{afiu} that such a connection also exists between the CY spaces, and 
that it corresponds to conifold transitions. These are the first examples 
of extremal transitions of exactly the type considered in \cite{gms} 
(i.e. abelian) for which the heterotic version is known. Recently this 
observation has been employed to enlarge the class of B and C models 
using toric methods \cite{canpera}.

\subsection{Exotic transitions and tensionless strings}

As mentioned in section 2.2, a strong coupling problem is encountered in 
$E_8\times E_8$ compactifications. The gauge coupling constant of the 
group coming from the $E_8$ with fewer instanton number diverges at the 
finite value of $\phi$ satisfying 
\beq
e^{-2\phi}=\frac{n}{2}
\label{dilastrong}
\eeq
Due to the $T$-dualities with $SO(32)$ vacua, the same behaviour is 
observed in these latter compactifications.

This phenomenon is also present in the F-theory framework, in which a 
geometrical interpretation can be given. The vev of the dilaton $\phi$ is 
related to the ratio of the K\"ahler classes $k_b$, $k_f$ of the base and 
fiber $\IP_1$'s in $\IF_n$, through the equation \cite{mv1}
\beq
e^{-2\phi}=\frac{k_b}{k_f}
\label{dilafth}
\eeq
One can determine that the size of the curve (divisor) $w_1=0$ is 
$k_b-n/2 k_f$, so that this cycle collapses for precisely the value 
(\ref{dilastrong}). The structure of central charges of the SUSY algebra 
implies \cite{sw6d} that a BPS saturated string becomes tensionless at 
that point, and such object is certainly found in F-theory, as a Type-IIB 
3-brane wrapped around the zero size cycle. The dynamics of these objects 
at the critical point is not infrared free, and the theory is at a 
non-trivial fixed point of the renormalization group. In some cases, this 
point lies at the intersection of different branches in moduli space, so 
that the tensionless string can mediate interesting phase transitions.

The physical behaviour at this point is encoded in the local geometry of 
the CY space near the vanishing curve. Actually, many features depend 
only on the geometry of the base, which once the curve has collapsed can 
be described as the projective space $\IP_2^{(1,1,n)}$. Since it only 
depends on the value of $n$, the results are valid for the 
different A, B, C and D families. They also apply to $SO(32)$ 
compactifications for the corresponding value of $n$.

For most values of $n$, not much is known about the theory at the critical 
point, since it is frozen at   a  Type-IIB   strong coupling regime 
\cite{witphase}. However the situation remains tractable for $n=0,1,2,4$, 
as we show in the following.

The case $n=0$ is special, since there is no gauge coupling divergence at 
finite $\phi$, and thus no singular behaviour. The case $n=2$, although 
seemingly presents such a divergence, can avoid the singular point in 
moduli space by turning on generic vevs for hypermultiplets. These issues 
are related to heterotic/heterotic duality and will be further explored 
in section 4.

The cases $n=1,4$ are interesting because the anomaly polynomial 
(\ref{polia}) becomes a perfect square, and can be cancelled by a GS 
mechanism using only the gravitational self-dual 2-form. Since the 
anti-selfdual 2-form in the tensor multiplet of the dilaton is not 
involved, a transition in which that tensor multiplet disappears is 
compatible with anomaly cancellation \cite{sw6d}. The new 
branch emerging from the transition point is parametrized by vevs for new 
hypermultiplets, and so is a Higgs branch. The geometric version of this 
argument is that the base spaces $\IF_1$, $\IF_4$ can be deformed to 
$\IP_2$. Since $h_{11}(\IP_2)=1$, F-theory compactifications on elliptic 
fibrations over $\IP_2$ have no tensors and provide a description for the 
Higgs branch.

The deformation $\IF_1 \to \IP_2$ is particularly simple to analyze, 
since $\IP_2^{(1,1,n)}$ is not singular for $n=1$ and the collapse of the 
curve $w_1=0$ is simply a smooth blowing down \footnote{This process 
is locally identical 
to the shrinking of a small $E_8$ instanton, mentioned in the transitions 
$\IF_n \to \IF_{n\pm1}$ in section 3.3.2}. Indeed, this simply amounts to 
setting $w_1=1$ in the fibration equations of the type (\ref{eq:fibr1}), 
yielding
\beq
y^2 = x^3 + {\tilde f} (z_1,z_2,w_2) x + {\tilde g} (z_1,z_2,w_2)
\eeq  
(and similar expressions for the B, C and D, or $SO(32)$ cases). Since 
$w_1$ no longer appears in the equations, new polynomial deformations are 
possible , reflecting the fact that the dilaton tensor multiplet 
has transformed into new hypermultiplets in order to preserve the 
cancellation of gravitational anomalies. Explicit counting of these 
deformations shows the Hodge numbers of the CY space change as
\beqa
\Delta (h_{11}) & = & -1 \nonumber\\
\Delta (h_{12}) & = & c_d-1
\eeqa
where $c_d$ is the Coxeter number of $E_d$ and $d=8,7,6$ and 5 for models
A, B, C and D. The groups $E_d$ do enter in the heterotic picture as
follows. Notice that for $n=1$, complete Higgsing of the non-Abelian
groups is possible in all models and this can be achieved by instantons of
$E_d \times U(1)^{8-d}$ that leave $U(1)^{8-d}$ unbroken in each $E_8$
(before further breaking to the diagonal combinations). In fact, the
transition to $n_T=0$ occurs when $k_2 \to k_2 + 1$, where $k_2$ corresponds
to an $E_d$ instanton. In the F-theory picture, the $E_d$ groups appear
because when the 2-cycle collapses in $\IF_1 \to \IP_2$, a 4-cycle of del 
Pezzo type shrinks in the CY \cite{mv2}, and there is a natural action of 
the Weyl group of $E_d$ on the 2-cycles inside this complex surface. It 
can be checked that the del Pezzo surfaces obtained for the A, B, C and D 
fibrations are of the correct $E_d$ type \cite{afiu}.

Existence of a Higgs branch with no tensor multiplets is also expected in 
the $n=4$ case. The Hodge numbers of the CY spaces change in the 
transition as
\beqa
\Delta(h_{11}) & = & -4 \nonumber \\
\Delta(h_{12}) & = & 1
\label{coxetern4}
\eeqa
which can be understood as follows. As the tensor multiplet disappears,  
anomaly cancellation conditions force the appearance
of 29 new hypermultiplets, 28 of which are employed in Higgsing the $SO(8)$
gauge symmetry and one remains in the final spectrum, providing for the 
increase in $h_{21}$. Also
4 Cartan generators are lost, thus explaining the change in $h_{11}$. 
This process is actually the second transition shown in 
eq.(\ref{transia}), where it was shown to be compatible with anomaly 
cancellation. We stress that in all the $n=4$ compactifications we have 
mentioned there is a terminal $SO(8)$, so the transition to the Higgs 
branch is allowed in all cases (on the F-theory side, the existence
of the corresponding $D_4$ singularity is discussed in \cite {mv2}).

\section{Heterotic/Heterotic Duality}

\subsection{Self-Dual $D=6$, $N=1$ Heterotic Vacua}

Heterotic/heterotic duality  in $D=6$, $N=1$ 
was first conjectured in
refs.~
\cite{Dufflu, Duffkhuri, Minasian, Duff} motivated by
heterotic/five-brane duality \cite{strom} in $D=10$. 
The  duality would involve an equivalence between the
strongly coupled heterotic compactified on K3   and the
weakly coupled string obtained by  wrapping  four 
of the dimensions of the the five-brane on K3.  
The duality between the corresponding 
gravitational  bosonic massless fields is given by the dictionary:
\beqa
\phi \  & \longleftrightarrow  &\ {\tilde \phi }=\ -\phi  \\
G_{MN} \  & \longleftrightarrow  &\ {\tilde {G}}_{MN} =\  e^{-\phi } G_{MN} \\
H  \  & \longleftrightarrow  &\ {\tilde  H  }=\  e^{-\phi} \*  H 
\label{dualbos} 
\eeqa
which shows that indeed this is a strong-weak coupling duality.

One of the most intuitive hints \cite{Minasian} for the existence
of a $D=6$  heterotic/heterotic duality is the way in which  the
anomaly
eight-form $I_8$ factorizes into the product of four-forms in $D=6$.
In fact, $I_8=X_4{\tilde X}_4$, with
\beqa
X_4\ & =& \frac{1}{4(2\pi )^2} \left( \tr R^2\ -\ V_{\alpha
} \tr F_{\alpha
}^2\right) \\
{\tilde X}_4\ & =& \frac{1}{4(2\pi )^2} \left ( \tr R^2\ -\ {\tilde
V}_{\alpha } \tr F_{\alpha }^2 \right) \quad ,\nonumber
\label{lasx}
\eeqa
where ${\alpha} $ runs over the gauge groups in the model.
This very symmetric form of $I_8$  suggests a duality under which
one exchanges the tree-level Chern-Simons contribution to the
Bianchi identity $dH={\alpha }'(2\pi )^2X_4$ with the one-loop
Green-Schwarz corrections to the field equations
$d{\tilde H} ={\alpha }'(2\pi )^2{\tilde X}_4$,  in agreement 
with  eq.(\ref{dualbos}) .  The kinetic term for the gauge 
bosons in eq.(\ref{sagno})   also show a potential 
duality under the exchange  $\phi \leftrightarrow -\phi $
as long  as ${\tilde V}_{\alpha }=V_{\alpha }$.
As we discussed in chapter 1, in these expressions
$V_{\alpha }$  is a (positive) tree-level coefficient which is
essentially the
Kac-Moody  level. On the other hand the coefficients ${\tilde
V}_{\alpha }$
are  associated to the Green-Schwarz mechanism, depend on the
massless spectrum of the model and they can be positive, negative or
zero.  It is thus clear that only certain  $D=6$ compactifications, those for
which ${\tilde V}_{\alpha}=V_{\alpha}$ have a  priori hope for presenting  
heterotic/heterotic duality.

In fact it is easy to obtain heterotic compactifications with ${\tilde V}_{\alpha }
=V_{\alpha }$.  In particular  we already obtained such an example in 
chapter 2, eq.(\ref{acuatro}). It is obtained  by a heterotic $Spin(32)/Z_2$ 
compactification on K3 with a certain $U(1)$ bundle  corresponding to
the $U(1)$ decomposition $SU(16)\times U(1)\in SO(32)$.  Using index
theorems one finds, for instanton number 24, hypermultiplets transforming as
$2( {\bf 120}) + 2( {\bf \overline{120}} )+ 20 ({\bf 1})$.  Using the formulae we 
gave in chapter two one easily finds $V_{SU(16)}={\tilde V}_{SU(16)}=2$.
Thus this provides the simplest example of heterotic/heterotic duality
within  the $Spin(32)/Z_2$ theory.
There are also perturbative $E_8\times E_8$  vacua  which display 
heterotic/heterotic duality.  Looking at eqs.(\ref{vtfor}), (\ref{sagnot})  
one sees that $E_8\times E_8$ vacua with $n=2$  (which corresponds to 
instanton numbers  $(k_1,k_2)=(14,10)$)  yields  duality for the  gauge group
in the first $E_8$.  Then the second $E_8$ is not self-dual but,
as remarked in ref.\cite{afiq2}, for generic
regions in the hypermultiplet moduli space the second $E_8$ is completely 
Higgsed away.

In fact the authors in refs.\cite{Dufflu, Duffkhuri, Minasian, Duff}   did not realized 
the existence of these {\it perturbative}  realizations of heterotic/heterotic
duality.  Chronologically 
the first explicit realization of  heterotic/heterotic duality in $D=6$, $N=1$
was the  {\it non-perturbative }  duality of ref.\cite{dmw} . 
These authors noticed that  $E_8\times E_8$ 
 compactifications with ${\tilde V}_{\alpha}=0$
 present  non-perturbative heterotic/heterotic duality.
Since this possibility is obviously not symmetric
under the exchange of $V_{\alpha }$ and ${\tilde V}_{\alpha}$,
it requires the dual gauge group to be generated by
non-perturbative (small instanton) effects as suggested in
\cite{witsm} . We already mentioned   in chapter 2
that   vacua with $n=0$  are T-dual to 
$Spin(32)/Z_2$ compactifications without vector 
structure  and we know that  such theories give rise
to non-perturbative enhanced gauge groups 
when instantons collapse to zero size.
This hypothesis is consistent with the fact that the 
 gauge groups generated by these non-perturbative effects
verify $V_{\alpha }=0$ (unlike the perturbative ones, which
obviously  have $V_{\alpha}>0$). The proposal can be
justified \cite{dmw}
by considering this duality as arising from two (dual) ways of
looking
at the compactification of the $D=11$ M-theory on $K3\times
S^1/Z_2$.
The two dual $D=6$ theories correspond to $E_8\times E_8$
heterotic $n=0$ compactifications on $K3$.

In fact  both perturbative and non-perturbative realizations of heterotic/heterotic
duality are   two aspects of a single self-dual theory, as can
be seen both in terms of Type-IIB orientifolds and F-theory.

\subsection{Heterotic/Heterotic Duality and Type-IIB Orientifolds}

Type-IIB , $Z_N$ orientifolds  \cite{hor} 
are obtained by compactifying  this string on 
$T^4$ and 
further  modding by $\{  \Omega, Z_N\}$  where $\Omega $ is the worldsheet
parity operation which acts like $\Omega z={\bar z}$ on the world-sheet 
complex coordinate
\cite{bs, bso, gp, dabol1, gj}.   $Z_N$ acts on $T^4$ in the same way described in 
chapter 2 for heterotic orbifolds.  The resulting vacua have $N=1$ SUSY in
$D=6$. The $\Omega$-twisted sectors are open strings, and 
consistency requires in general the 
presence of two types of  boundaries: nine-branes an five-branes.  Tadpole cancellation
constraints the number of  those as well as restricting the embedding of the
$Z_N$ symmetry on the nine-brane and five-brane Chan-Paton (CP) factors.  We will be 
interested in the particular case of $Z_2$ which corresponds to the
Bianchi-Sagnotti-Gimon-Polchinski (BSGP)  class of models
\cite{bso, gp}. Notice that
$T^4/Z_2$ corresponds to an orbifold limit of K3 with 16  $Z_2$ fixed points.
From the closed string sector of the orientifold one gets the usual gravity plus 
tensor multiplet. In addition, one gets 4 moduli from the untwisted closed string and 16 
more from the twisted ones.   Tadpole cancellation requires the presence of 32
nine-branes (usual  Type-I open string boundaries) and 32 
(Dirichlet) five-branes .
The model with maximal gauge symmetry is 
obtained for a configuration  with no Wilson lines on the nine-branes and 
all five-branes sitting on the same fixed point.  In this case the
gauge group is $U(16)_{99}\times U(16)_{55}$, which corresponds to open 
strings stretching between a couple of  nine(five)-branes respectively. Those 
strings also give rise to hypermultiplets transforming like
$2({\bf 120},{\bf 1})_{99}+2({\bf 1},{\bf 120})_{55}$. Open strings 
stretching  between 
nine-branes and five-branes give rise to extra massless hypermultiplets
transforming like $({\bf 16},{\bf 16})_{59}$.  The reader may check that 
this massless spectrum indeed is anomaly free.   In Type-I  open strings,
T-duality exchanges Neumann and Dirichlet boundary conditions and 
in the present case this means the exchange of  five-branes and nine-branes.
Indeed, this particular configuration is explicitly invariant 
under T-duality.  The $D=6$  anomaly polynomial for this model was 
computed in ref.\cite{berkooz}  and found to have the form:
\beq
A_8\ =\ (R^2\ -\ 2 F^2_9 )(R^2\ -\  2 F^2_5)  \ . 
\label{a8gp}
\eeq
Comparing this to eq.(\ref{factor})  one finds  for  $U(16)_{99}$   $V_9=2$
$,{\tilde V}_{9}=0$  whereas for $U(16)_{55}$  one has 
$V_5=0$ and  ${\tilde V}_5=2$.  As remarked in \cite{berkooz} , 
looked from the dual {\it heterotic } point  of view one would say that the first
$U(16)$ is a perturbative gauge group with ${\tilde V}=0$, similar to the
$n=0$ $E_8\times E_8$ vacua or  the T-dual  $Spin(32)/Z_2$
models without vector structure.  However, the second $U(16)$  has
non-perturbative origin as indicated by the fact that $V=0$ for that 
gauge group.  From the Type-I  point of view, T-duality 
exchanges  both $U(16)$'s. Thus the non-perturbative
heterotic/heterotic duality  maps to usual T-duality on the Type-I dual side.

In this setting is easy to see the connection   between  the $n=2$ and
$n=0$ realizations of heterotic/heterotic duality. Consider the above
model with $U(16)^2$ symmetry. If we give a non-vanishing
vev  to the hypermultiplets in  $({\bf 16}, {\bf 16})_{59}$  the gauge 
group is broken to the diagonal  one $U(16)_{diag}$  and one has
hypermultiplets transforming like $4({\bf 120})+20 ({\bf 1})$. But this is 
precisely the massless  spectrum of the $SO(32)$ perturbative  model 
that we showed in the previous subsection.  We also have 
$V_{diag}=V_9$ $+V_5=2+0=2$  and ${\tilde V}_{diag}={\tilde V}_9$ $+{\tilde 
V}_5=0+2=2$
so that indeed ${\tilde V}_{diag}/V_{diag}=1$ as required to have self-duality.

\subsection{Heterotic/Heterotic Duality and F-theory}

The F-theory description of heterotic/heterotic duality for $n=0$ turns 
out to be particularly simple. Since $\IF_0=\IP_1\times \IP_1$, there is 
a natural symmetry in the moduli space which interchanges both $\IP_1$'s. 
By eq.(\ref{dilafth}) this implies the change $\phi \to -\phi$, 
suggesting the identification of this action with heterotic/heterotic 
duality. Observe moreover that perturbative symmetries (associated to 
singular elliptic fibers over the `base' $\IP_1$) are mapped to 
non-perturbative ones (corresponding to singular elliptic fibers over the 
`fiber' $\IP_1$), thus explaining the non-trivial action of the duality on 
the hypermultiplets \cite{dmw}.

There is also an explanation for the existence of heterotic/heterotic 
duality for $n=2$. As remarked above, the naive strong coupling 
singularity in the gauge factor with fewer instantons can be avoided by 
Higgsing that gauge symmetry, i.e. by turning on vevs for 
hypermultiplets. In the F-theory language this must be realized as a 
complex structure deformation of the fibration over $\IF_2$, e.g. 
$\IP_4^{(1,1,2,8,12)}[24]$. From the 243 deformations of this CY, 242 can 
be represented as polynomial deformations, and do not change the singular 
geometry on the base once the $w_1=0$ curve has been blown 
down, i.e. once the $\phi$-value (\ref{dilastrong}) has been reached. At 
that point the base becomes $\IP_2^{(1,1,2)}$, with an $A_1$ singularity.
However, as remarked 
in \cite{aghh,mv1} there is a non-polynomial deformation. For a generic 
value of this, the CY cannot be embedded in the projective space 
$\IP_4^{(1,1,2,12,18)}$. This in particular implies that when one reaches 
the problematic $\phi$ value, the base space is not $\IP_2^{(1,1,2)}$, but 
a deformation of it, with the $A_1$ singularity properly smoothed. Then 
no tensionless string is found unless this modulus is tuned to zero.

Actually, this complex structure deformation transforms $\IF_2$ into 
$\IF_0$, so that elliptic fibrations over both surfaces lead to identical 
CY's. Only when this parameter is turned off, i.e. at codimension one in 
the moduli space, precisely in the case that the CY can be embedded in 
$\IP_4^{1,1,2,12,18}$, one meets the strong coupling singularity. The 
possibility of avoiding the singularity is also present in the physics of the
non-critical string that appear at that point \cite{witphase}. Since the 
local geometry is hyperk\"ahler, the string carries a tensor as 
well as an extra hypermultiplet degree of freedom. Its tension depends on 
all five parameters, and only vanishes when the scalar in the tensor 
multiplet {\em and} those in the hypermultiplet are properly tuned. For 
generic vevs for this hypermultiplet, the singularity is circumvented.

The results of section 4.2 concerning the equivalence of $n=0$ and $n=2$ 
compactifications in models with enhanced symmetries can also be 
reproduced in F-theory. As shown in ref.\cite{bikmsv} one can construct $n=0$
compactifications with perturbative and non-perturbative enhanced gauge 
groups by forcing singular elliptic fibers over the `base' and `fiber' 
$\IP_1$'s in $\IF_0$ (the particular configuration $U(16)\times U(16)$ 
discussed in section 4.2 has been explicitly worked out in 
\cite{aspfz2}). Upon a complex structure deformation, precisely the 
non-polynomial one mentioned above, the two curves merge 
smoothly at their intersection point to form a single curve of singular 
fibers, representing the diagonal subgroup of the original symmetry 
\cite{bikmsv}. 
The final configuration is manifestly self-dual, and corresponds to a 
perturbative enhanced model from the $n=2$ viewpoint.

\subsection{ $D=4$, $N=2$  Heterotic/Heterotic Duality}

Let us now consider the  $D=4$ heterotic models obtained upon further
compactification of the above $D=6$  heterotic duals  on a 2-torus.
This case was considered in ref.~\cite{Duff} and briefly mentioned in
ref.~\cite{dmw}. The resulting
$N=2$ theory has the usual toroidal vector multiplets
$S,T,U$ related to the coupling constant and the size and shape of
the 2-torus. When the $D=6$ theory is
dimensionally reduced to four dimensions,
the underlying duality exchanges the roles of
$S$ and $T$ \cite{Duff, Pierre}. Including mirror symmetry on the
torus,
one thus expects complete  $S-T-U$  symmetry in this type of vacua
\cite{Duff, dmw, Rahmfeld, klm, Duffrev, cclmr}.
Thus, on top of the usual perturbative $SL(2,\IZ)_T\times
SL(2,\IZ)_U$
dualities, a non-perturbative $SL(2,\IZ)_S$  $S$-duality \cite{filq}
is expected. This $N=2$ model has the toroidal $U(1)^4$ as
generic gauge group, and as matter, $244$ neutral
hypermultiplets (it corresponds to the heterotic construction of
model $B$
of ref.~\cite{kv}). At particular points in moduli space,
enhanced gauge groups such as $E_7\times E_7$ can appear.

A natural question is the following: What is the $D=4$ equivalent of
the ${\tilde V}_{\alpha }=0$  or  ${\tilde V}_{\alpha }=V_{\alpha }$
conditions we had in $D=6$  in order to
have heterotic/heterotic duality? It turns out that the equivalent
condition
in $D=4$ can be phrased as a condition on the $N=2$ $\beta
$-functions
of the gauge groups present at enhanced points in moduli space.
Indeed, the $N=2$, $D=4$ $\beta $-function of a given
gauge factor can be written in terms of
the corresponding $D=6$ ${\tilde V}_{\alpha }$ coefficient.
More explicitly one finds \cite{afiq2}
\beq
\beta_\alpha^{N=2} \ = \ 12\left(1 +  \frac{\tilde
V_\alpha}{V_\alpha}\right) \quad .
\label{betagen}
\eeq
Thus, the condition to get heterotic/heterotic duality in  $N=2$,
$D=4$ models reads
\beqa
\beta_\alpha^{N=2} &=&  12 \quad\quad  \hbox{(symmetric  $E_8\times
E_8$
embeddings)} \nonumber \\
\beta_\alpha^{N=2} &=&  24 \quad\quad \hbox{(non-symmetric
$E_8\times E_8$ embeddings)} \quad .
\label{dos}
\eeqa
Notice that in both cases the $N=2$ models are
non-asymptotically free.
In the first case ($\beta_{\alpha}$ =\, 12),
consistently with the DMW hypothesis in $D=6$, there
should be points in moduli space in which new gauge groups
of a non-perturbative origin should appear.
Those are required to obtain full duality.  In the second case
($\beta_{\alpha}$ =\, 24) this is not expected but explicit
duality should be apparent.

One can think of the following  consistency check . 
We know the form of the holomorphic
$N=2$ gauge kinetic function $f_{\alpha }$ for the gauge groups
inherited from
$E_8\times E_8$.  For a $K3\times T^2$ compactification of the type
discussed here one has \cite{Kaplouis, deWit}
\beq
f_{\alpha }\ =\ k_{\alpha }S_{inv}\ - \
\frac{{\beta }^{N=2}_{\alpha }}{4\pi } \log \left(\eta (T)\eta
(U)\right)^4 \quad ,
\label{efe}
\eeq
where $\eta $ is the Dedekind function and $S_{inv} $ is given by  :
\beq
S_{inv}\equiv S-\frac{1}{2}\partial_T\partial_U
h^{(1)}(T,U)-\frac{1}{2\pi}\log \left(J(T)-J(U)\right)
+{\rm const}.
\label{sinv}
\eeq
Here $h^{(1)}(T,U)$  is the moduli-dependent one-loop correction to
the $N=2$ prepotential ${\cal F}$  and $J$ is the
absolutely modular invariant function. 
Now, we know that the large-$T$ limit of $f_\alpha$ must reproduce
the result in eq.~(\ref{sagno}).
It is not clear that this
follows from  the above equations. 
However one can check   \cite{afiq2}  that   for large $T$ one has
$S_{inv}\rightarrow S-T$  so that one gets
 for $k_\alpha=1$
\beq
\lim_{T\rightarrow \infty } f_{\alpha } \ =\
S\  +T \left(\frac{\beta_{\alpha }^{N=2}}{12}\ -\ 1\right)\ =\
S\ +\  \frac{{\tilde V}_{\alpha }}{V_{\alpha }} \ T \quad ,
\label{limites}
\eeq
which is just the $D=4$ version of formula (\ref{sagno}).
We thus see that if any of the $\beta_\alpha^{N=2}$ is smaller
than  12, the large-$T$ limit
gives rise to gauge kinetic terms of the wrong sign, 
which is just a four dimensional reflection of the 
$D=6$ singularity be discussed previously.
Notice the different large-$T$ behaviour of the
two heterotic/heterotic dualities under consideration.
In the one  proposed  in \cite{dmw}   one has ${\tilde V}_{\alpha }=0$
and $f\rightarrow S$. In the alternative  $n=2$  case  \cite{afiq2} , 
one has $f\rightarrow S+T$, a $S\leftrightarrow T$
invariant result.

\vskip0.6cm

\centerline{\bf Acknowledgements}
L.E.I  thanks the organizers  
of the  APCTP  Winter School on Duality 
for a very stimulating meeting.  Most of the original material in the
lectures was worked out in collaboration  with 
G. Aldazabal, A. Font, F. Quevedo and  G. Violero. 
Our gratitude goes to them for a very enjoyable  collaboration.

\end{document}